\documentclass[a4paper,12 pt]{article}

\usepackage{graphicx} 
\usepackage{amsmath, amssymb, amsfonts, amsthm}
\usepackage{multicol,multirow}
\usepackage{bm}
\usepackage{xcolor}
\usepackage{appendix}
\usepackage{url}
\usepackage{longtable}
\usepackage[margin=1 in]{geometry}
\usepackage[authoryear]{natbib}
\usepackage{authblk}

\newcommand{\btheta}{ \mbox{\boldmath $\theta$}}
\newcommand{\bmu}{ \mbox{\boldmath $\mu$}}

\newcommand{\bbeta}{ \mbox{\boldmath $\beta$}}

\newcommand{\bSigma}{ \mbox{\boldmath $\Sigma$}}

\newcommand{\bzero}{ \mbox{\boldmath $0$}}
\newcommand{\bA}{ \mbox{\bf A}}

\newcommand{\bX}{ \mbox{\bf X}}

\newcommand{\bZ}{ \mbox{\bf Z}}
\newcommand{\by}{ \mbox{\bf y}}
\newcommand{\bY}{ \mbox{\bf Y}}

\newcommand{\bs}{ \mbox{\bf s}}

\newcommand{\bW}{ \mbox{\bf W}}

\newcommand{\bQ}{ \mbox{\bf Q}}
\newcommand{\bI}{ \mbox{\bf I}}

\newcommand{\iid}{\stackrel{iid}{\sim}}
\newcommand{\indep}{\stackrel{indep}{\sim}}

\newcommand{\calD}{{\cal D}}
\newcommand{\calS}{{\cal S}}

\newcommand{\bbi}{\mbox{$\mathbb{I}$}}

\newcommand{\beq}{ \begin{equation}}
\newcommand{\eeq}{ \end{equation}}
\newcommand{\beqn}{ \begin{eqnarray}}
\newcommand{\eeqn}{ \end{eqnarray}}

\newcommand{\beqs}{ \begin{equation*}}
\newcommand{\eeqs}{ \end{equation*}}
\newcommand{\beqns}{ \begin{eqnarray*}}
\newcommand{\eeqns}{ \end{eqnarray*}}

\newcommand{\rpm}{\sbox0{$1$}\sbox2{$\scriptstyle\pm$}
  \raise\dimexpr(\ht0-\ht2)/2\relax\box2 }

\def\Lp{\left(}
\def\Rp{\right)}
\def\LP{\left\{ } 
\def\RP{\right\}}

\title{Bayesian Variable Selection for Censored Spatial Responses with Application to PFAS Concentrations in California} 

\author[1]{Suman Majumder}
\affil[1]{Interdisciplinary Statistical Research Unit, Indian Statistical Institute, Kolkata, West Bengal, India}
\author[2]{Indranil Sahoo \footnote{Corresponding Author}}
\affil[2]{Department of Statistics, Virginia Commonwealth University, Richmond, Virginia, USA}
\date{ }

\begin{document}
\maketitle

\vspace{-30 pt}

\begin{abstract}
    Per- and polyfluoroalkyl substances (PFAS) are persistent environmental pollutants of major public health concern due to their resistance to degradation, widespread presence, and potential health risks. Analyzing PFAS in groundwater is challenging due to left-censoring and strong spatial dependence. Although PFAS levels are influenced by sociodemographic, industrial, and environmental factors, the relative importance of these drivers remains unclear, highlighting the need for robust statistical tools to identify key predictors from a large candidate set. We present a Bayesian hierarchical framework that integrates censoring into a spatial process model via approximate Gaussian processes and employs a global-local shrinkage prior for high-dimensional variable selection. We evaluate three post-selection strategies, namely, credible interval rules, shrinkage weight thresholds, and clustering-based inclusion and compare their performance in terms of predictive accuracy, censoring robustness, and variable selection stability through cross-validation. Applied to PFOS concentrations in California groundwater, the model identifies a concise, interpretable set of predictors, including demographic composition, industrial facility counts, proximity to airports, traffic density, and environmental features such as herbaceous cover and elevation. These findings demonstrate that the proposed approach delivers stable, interpretable inference in censored, spatial, high-dimensional contexts, thereby offering actionable insights into the environmental and industrial factors affecting PFAS concentrations.\\

\end{abstract}

\section{Introduction} \label{intro}

Per- and polyfluoroalkyl substances (PFAS), often called ``forever chemicals", are a class of synthetic compounds extensively used in industrial and consumer products, including firefighting foams, non-stick cookware, and industrial coatings, because of their persistence, resistance to degradation, and environmental mobility. However, these properties have also made PFAS a growing public health concern. They accumulate in groundwater and drinking water supplies, persist in human tissue, and have been linked to a range of adverse outcomes including immune suppression, thyroid disruption, liver damage, reproductive disorders, and cancer \citep{dewitt2015toxicological, costello2022exposure, dunder2023associations}. This public health threat has catalyzed increased regulatory attention and environmental monitoring, especially in regions such as California, where groundwater is a critical source of drinking water. Recognizing the severity of the issue, the United States Environmental Protection Agency (EPA) finalized its first legally enforceable maximum contaminant levels (MCLs) in drinking water for six PFAS in April 2024. The limits were set at $4.0$ parts per trillion (ppt; equivalent to nanograms per litre, $ng/L$) for perfluorooctanoic acid (PFOA) and perfluorooctane sulfonic acid (PFOS), and $10.0$ ppt for perfluorononanoic acid (PFNA), perfluorohexane sulfonic acid (PFHxS), and hexafluoropropylene oxide dimer acid (GenX), along with a hazard index limit for mixtures. The rule also requires that water systems complete monitoring by 2027 and implement remediation measures by 2029 \citep{read2024pfas}. This landmark regulation highlights the urgent need for reliable and scalable statistical methods that can accurately detect and quantify the environmental and demographic factors driving PFAS contamination.

Recent analyses of PFAS data span a wide range of statistical paradigms, reflecting diverse inferential goals and data structures. For left‐censored concentration measurements, several studies have applied standard or mixed-effects Tobit regression to handle values below reporting limits while estimating the effects of behavioral and demographic predictors \citep{mulhern2021longitudinal, barton2020sociodemographic}. \cite{sahoo2025computationally} proposed a fast and scalable Bayesian framework that simultaneously accounts for spatial dependence and substantial left-censoring to analyze PFAS concentrations across California. In exposure assessment and occurrence modeling, hierarchical Bayesian frameworks have been developed to estimate national patterns of occurrence in public drinking water systems with formal uncertainty quantification \citep{cadwallader2022bayesian}. Bayesian hierarchical models have also been applied to characterize human toxicokinetics of multiple PFAS from contaminated drinking water and to assess exposure disparities and health risks between private well and public-supply tap water in the United States \citep{chiu2022bayesian, smalling2023per}. In addition, machine-learning methods such as random forests and related classifiers and regressors have been used to predict contamination risk and PFAS levels from geospatial, source-proximity, and environmental predictors, primarily for screening and prioritization of sampling \citep{barton2025data, deluca2023using}. Finally, in epidemiological settings with correlated exposures, Bayesian Kernel Machine Regression have been used to assess potentially nonlinear, interactive effects of multiple PFAS on health outcomes \citep{marfo2025combined}.

Despite these advances, existing approaches face significant limitations. Methods that explicitly account for spatial dependence in PFAS data often neglect the role of meaningful covariates, rendering the resulting inference less informative for identifying environmental and sociodemographic drivers of contamination. Conversely, studies that investigate covariate effects on PFAS concentrations typically ignore spatial correlation and are not designed to accommodate high-dimensional, left-censored data. From a methodological standpoint, geostatistical models such as Gaussian Processes \citep{schulz2018tutorial} with Mat\'ern covariance structures provide the framework for many spatial analyses in environmental science \citep{genton2001classes, cressie2015statistics, sahoo2023estimating}. However, directly applying such frameworks to censored data at scale leads to computational intractability. Approaches such as Expectation-Maximization (EM) based data augmentation \citep{militino1999analyzing, ordonez2018geostatistical} or Monte Carlo approximations \citep{stein1992prediction, tadayon2017bayesian, sahoo2021contamination} have been employed to address censoring in spatial inference. However, these methods remain limited in scalability. The Stochastic Partial Differential Equation (SPDE) approach proposed in \cite{sahoo2025computationally} improves the computational speed by approximating Gaussian processes with Gaussian Markov random fields (GMRFs), enabling efficient Bayesian inference in large spatial domains. Nevertheless, few existing applications extend this paradigm to simultaneously integrate covariate-driven inference, censoring, high-dimensionality, and scalable computation in the context of PFAS contamination studies.

In this paper, we develop a computationally efficient hierarchical Bayesian framework that simultaneously addresses spatial dependence and heavy left-censoring while enabling systematic covariate selection in the study of PFAS concentrations. We integrate variable selection methods into the scalable SPDE–GMRF approximation of \citet{sahoo2025computationally}, thereby achieving a balance between computational scalability and interpretability while identifying key environmental, hydrological, and sociodemographic drivers of PFAS contamination (Section \ref{ss:meth_base}). Unlike the Gaussian priors induced on the regression coefficients in \cite{sahoo2025computationally}, which resemble ridge regression and impose uniform shrinkage without facilitating variable selection, we adopt a global–local shrinkage prior, specifically the horseshoe$+$ prior \citep{bhadra2017horseshoe+}, to effectively distinguish between significant and non-significant predictors (Section \ref{ss:meth_varsel}). To complement this framework, we evaluate three strategies for variable selection (Section \ref{sss:zerocoefmeth}). First, we utilize a posterior credible interval–based rule, which retains variables whose $95\%$ credible intervals exclude zero. Second, we implement a shrinkage-weight thresholding strategy \citep{bhadra2017horseshoe+}, which uses the posterior expectation of effective shrinkage weights to adaptively select predictors and is known to be effective for ultra-sparse settings. Finally, we implement a clustering-based method \citep{li2017variable, chapagain2024variable} which partitions coefficients into significant and non-significant groups via sequential two-means clustering, and works well in moderately sparse settings. Through carefully designed simulation studies, we evaluate these strategies in terms of predictive performance, robustness to censoring, and stability of variable inclusion (Section \ref{s:sim}). After some preprocessing and exploratory analysis (Section \ref{s:data_des}), we apply our proposed modeling framework to PFOS concentration data from California’s Groundwater Ambient Monitoring and Assessment (GAMA) program and identify a consistent set of influential covariates, thereby highlighting their implications for PFAS transport and persistence in groundwater in California (Section \ref{s:data_ana}). Finally, some concluding remarks are presented in Section \ref{s:conc}. 

\section{PFAS Data Description} \label{s:data_des}

This study focuses on California, where groundwater has been extensively sampled for PFAS \citep{george2021machine}. We restrict the study area to 40 counties in which at least 25\% of the population relies on groundwater for drinking, thereby emphasizing exposure through drinking water rather than PFAS concentration levels or chronic health incidence. This excludes major urban centers such as Los Angeles and the Bay
Area, where groundwater make only a minor contribution to drinking water. Groundwater PFAS concentrations are obtained from California’s Groundwater Ambient Monitoring and Assessment (GAMA) program for the period 2016 – 2022, with $99\%$ of samples collected after 2019. PFAS concentrations are measured in nanograms per litre ($ng/L$), with values below the detection limit treated as censored. The minimum detection limits also vary by testing method and across sites. Among all PFAS species measured, perfluorooctane sulfonate (PFOS) consistently exhibits the highest concentrations at most sites, and therefore we focus our analysis on PFOS as the contaminant of interest in this study.

\begin{figure}
    \centering
    \includegraphics[width=0.45\linewidth,trim= 0 15 15 0,clip]{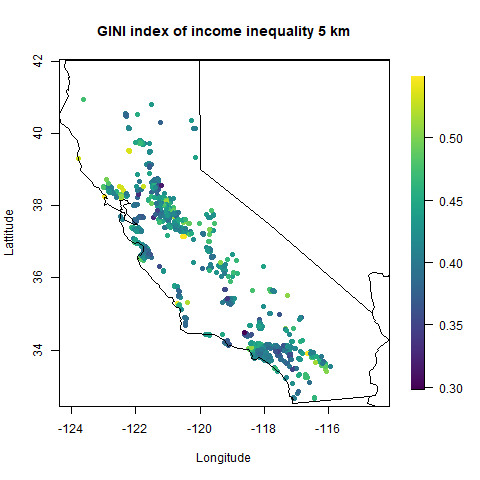} \includegraphics[width=0.45\linewidth,trim= 0 15 15 0,clip]{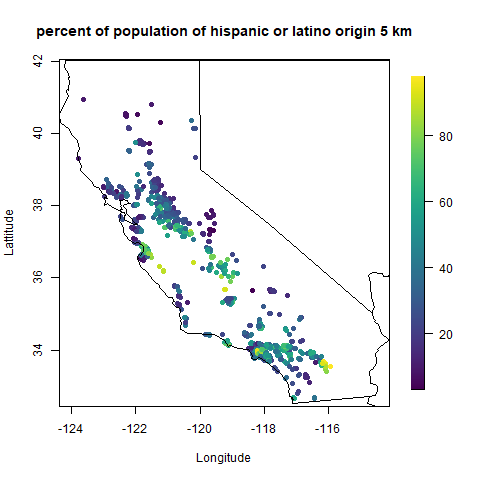} \includegraphics[width=0.45\linewidth,trim= 0 15 15 0,clip]{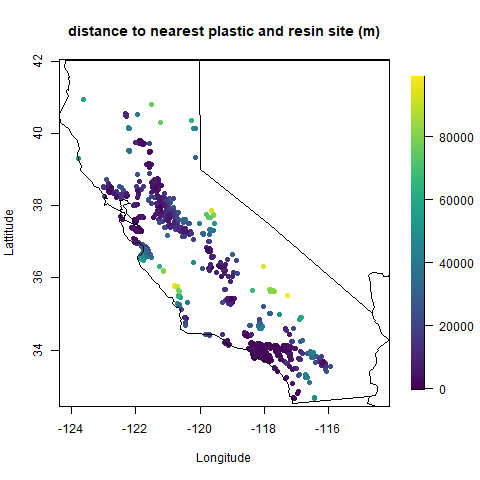} \includegraphics[width=0.45\linewidth,trim= 0 15 15 0,clip]{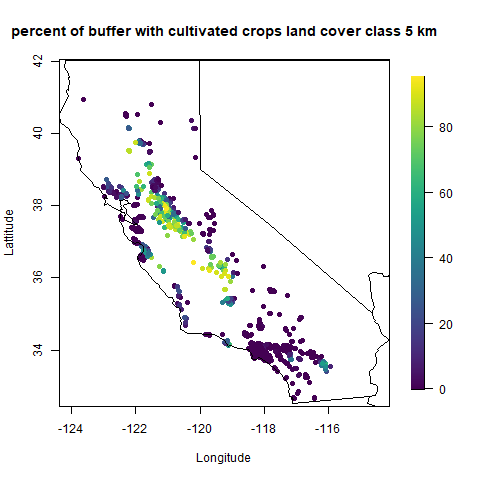} \includegraphics[width=0.45\linewidth,trim= 0 15 15 0,clip]{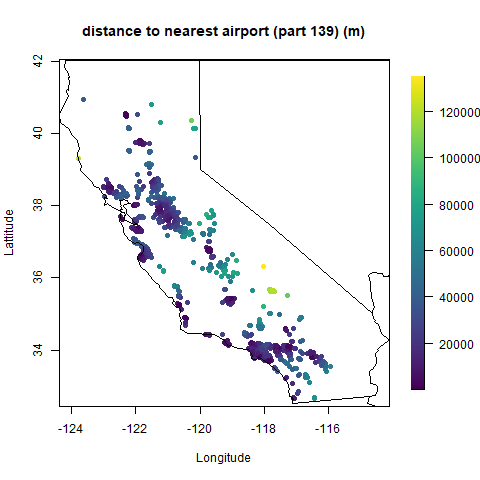} \includegraphics[width=0.45\linewidth,trim= 0 15 15 0,clip]{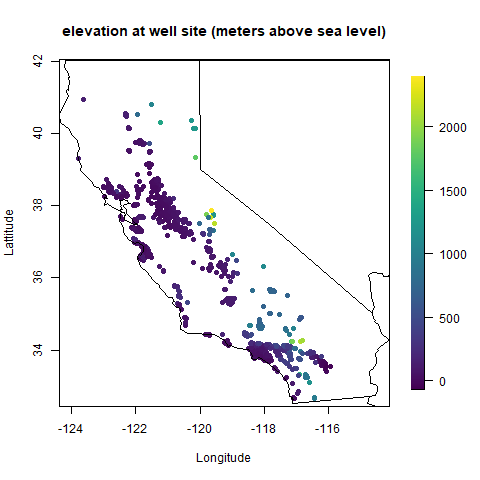}
    \caption{Spatial distributions of selected covariates included in the study.}
    \label{fig:eda_cov}
\end{figure}

Based on empirical knowledge of PFAS sources and environmental transport, our covariates include sociodemographic characteristics, potential industrial sources of PFAS, land cover attributes, and the incidence of chronic health conditions. Brief descriptions of these covariates are provided below. To avoid prematurely excluding relevant predictors, we adopt a broad set of candidate variables, allowing the subsequent Bayesian variable selection framework to identify the most parsimonious set of drivers. Spatial descriptive maps of a few selected covariates are presented in Figure \ref{fig:eda_cov}. A complete list of covariates used in this study has been provided in Table \ref{tab:app_covdes} in \ref{app2}.

\subsection*{Sociodemographic Characteristics}

Population characteristics including age, race, and income were obtained at the ZCTA and census tract levels in California from the U.S. Census Bureau’s 5-year American Community Survey (ACS-5) for 2015 (\url{https://www.census.gov/data/developers/data-sets/acs-5year.html}; accessed August 27, 2025). Additional sociodemographic data were retrieved from California’s Office of Environmental Health Hazard Assessment’s CalEnviroScreen 4.0 dataset, an interactive mapping tool that compiles environmental, socioeconomic, and public health data (\url{https://oehha.ca.gov/calenviroscreen}; accessed August 27, 2025). The CalEnviroScreen data, available at the census tract level, were linked to individual sites and then averaged across census tracts within each ZCTA.

\subsection*{PFAS Sources}

Potential industrial sources of PFAS such as airports, fire training facilities, and chemical manufacturing sites (see Table \ref{tab:app_industries} in \ref{app2}) were identified from the EPA’s PFAS Analytics Tools and quantified within 1- and 5-km buffers around sites as well as within ZCTAs (\url{https://awsedap.epa.gov/public/extensions/PFAS_Tools/PFAS_Tools.html}; accessed August 27, 2025). In addition, distances between each site and the nearest PFAS industry source were calculated. Both the number of PFAS-related facilities and their proximity to sites were included as predictors of exposure.

\subsection*{Landcover and Environmental Data}

Land use and landcover information were obtained from the 2016 National Land Cover Dataset (NLCD) provided by the Multi-Resolution Land Characteristics Consortium \citep{wickham2021thematic, dewitz2019national}. The dataset classifies landcover into 16 categories at 30 m resolution across the United States. The NLCD Land Cover Change Index, measuring landcover changes from 2001 – 2019, was also incorporated in the study. Landcover data were summarized for each ZCTA and within 1- and 5-km buffers around sites. Soil properties, including percent organic matter, clay content, and pH, were obtained from the POLARIS database \citep{chaney2016polaris, moro2021xpolaris}. Site elevations (meters above sea level) were extracted from the United States Geological Survey using the \texttt{elevatr} package \citep{hollister2017elevatr} in \texttt{R}.



\subsection{Data Preprocessing and Exploratory Analysis}

The original GAMA dataset contained measurements of PFOS concentrations at 22,719 observations across California. However, covariate information was only obtained at 2,776 locations in the study area. Since PFOS measurements were not available at 36 of these locations, we restricted our analysis to the remaining 2,740 locations where both PFOS concentrations and covariates were available.

Among the 289 candidate covariates, categorical variables such as Race, Gender, Home ownership status and Land cover type were dummy-coded with baseline categories (`male', `white', and `water-covered land'), and the corresponding columns were omitted from the design matrix to avoid collinearity. In addition, certain covariates were derivatives of each other; for example, the pollution burden score and its corresponding percentile. In such cases, the percentile-based covariates were excluded, as they did not contribute additional information and made the design matrix ill-conditioned. Another example was the covariate total industry, which was computed as the sum of several other covariates representing counts of different industry types in the vicinity of each site. Since its inclusion would introduce perfect collinearity, we excluded it in favor of the disaggregated covariates. A few additional covariates contained only zero entries, and were also omitted.

The dataset also included medical outcome variables, such as the age-adjusted rate of emergency hospital visits due to asthma or cardiovascular diseases, as well as low birth rates. Although these variables may be associated with PFOS concentrations, they are possibly downstream consequences of it, rather than predictors. Including them could, in principle, capture statistical associations, but their interpretation as covariates in this context would be erroneous, and they were therefore excluded as well. After these exclusions, 261 of the 289 covariates remained in the analysis. Following the removal of missing data in both covariates and responses, observations at 2,394 locations in the study area were retained. Out of these, PFOS concentrations were left-censored at 1,644 (68.67\%) locations.

\begin{figure}
    \centering
    \includegraphics[width=0.49\linewidth]{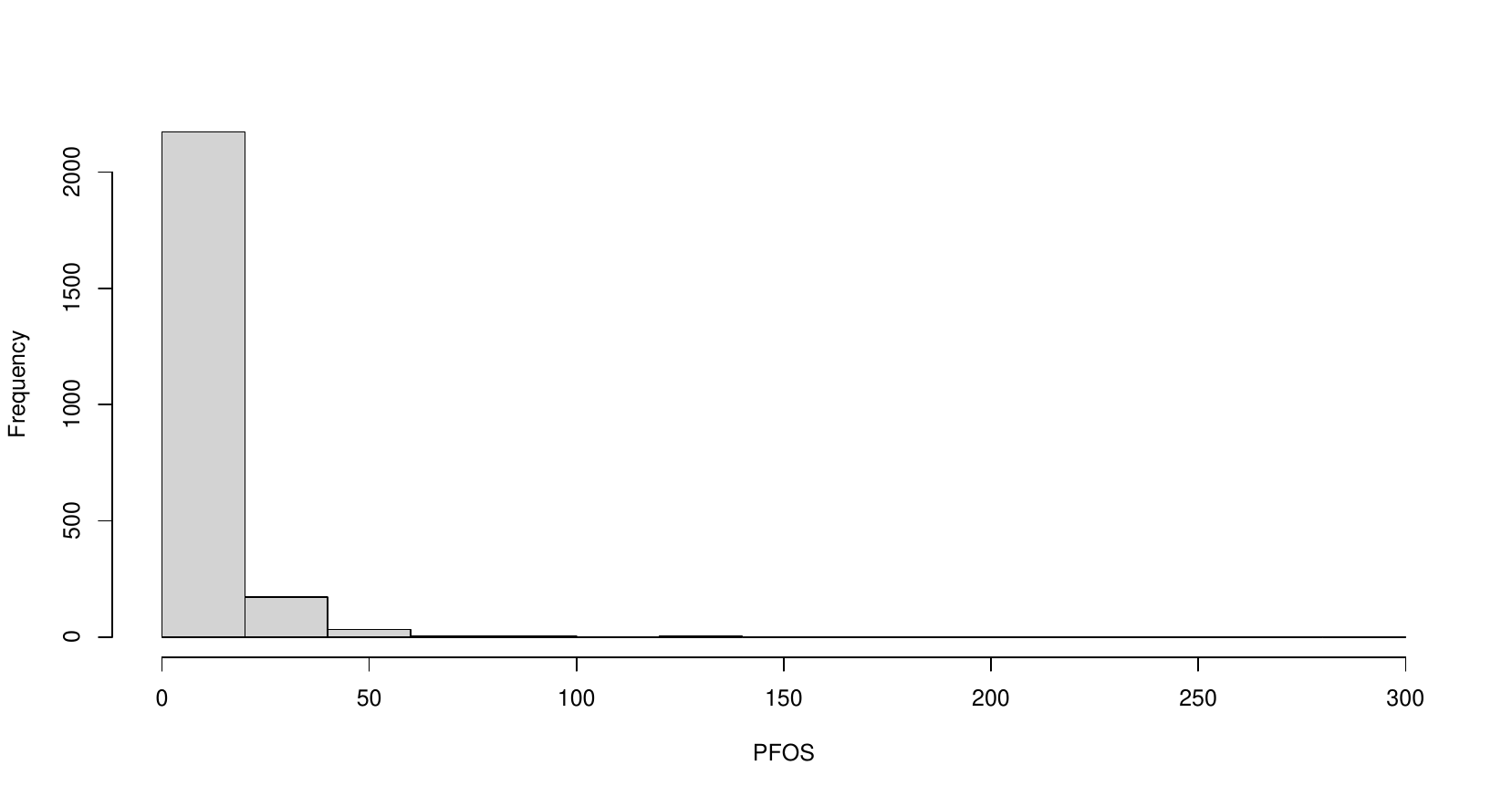} \includegraphics[width=0.49\linewidth]{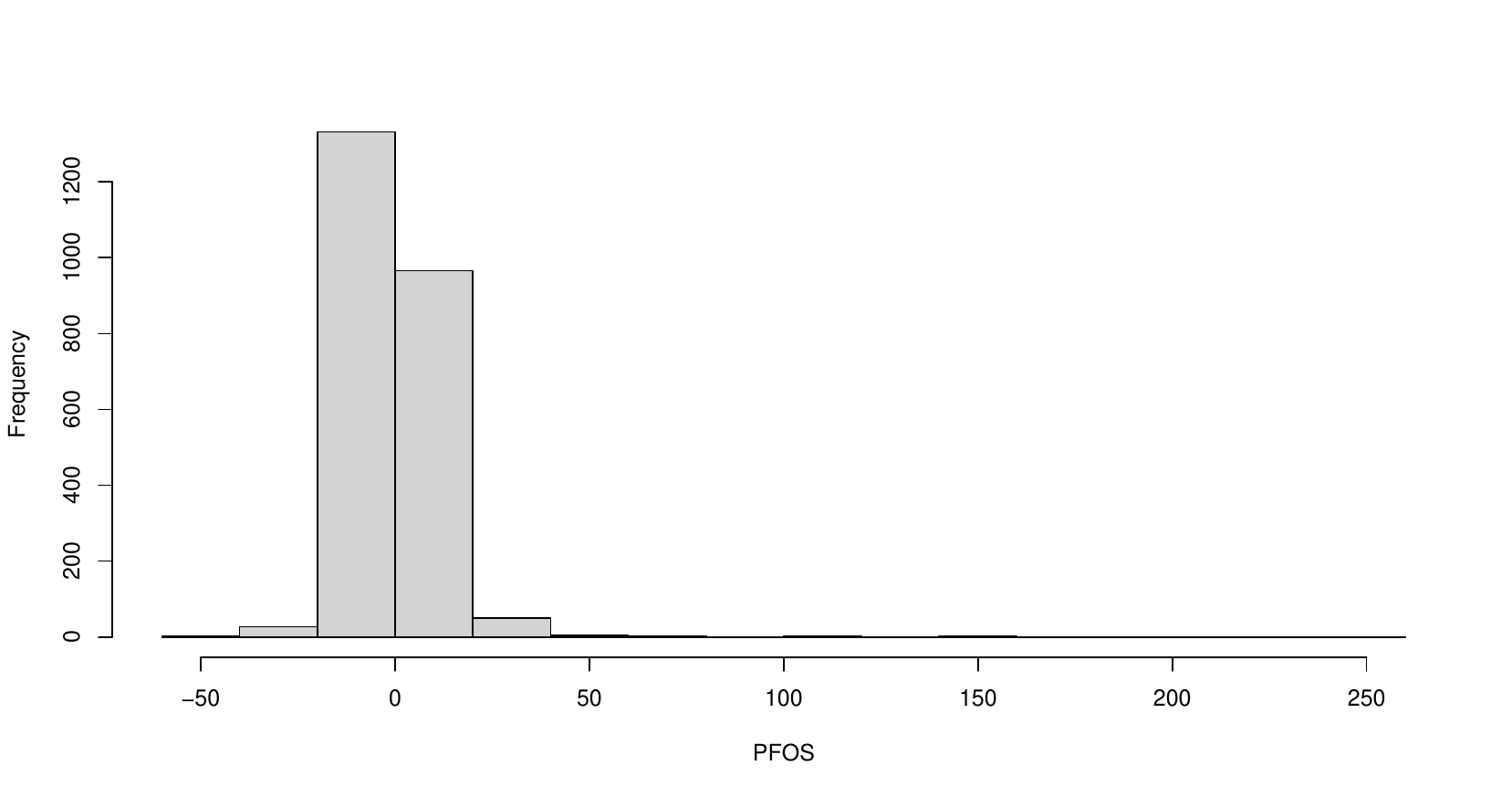}
    \includegraphics[width=0.49\linewidth]{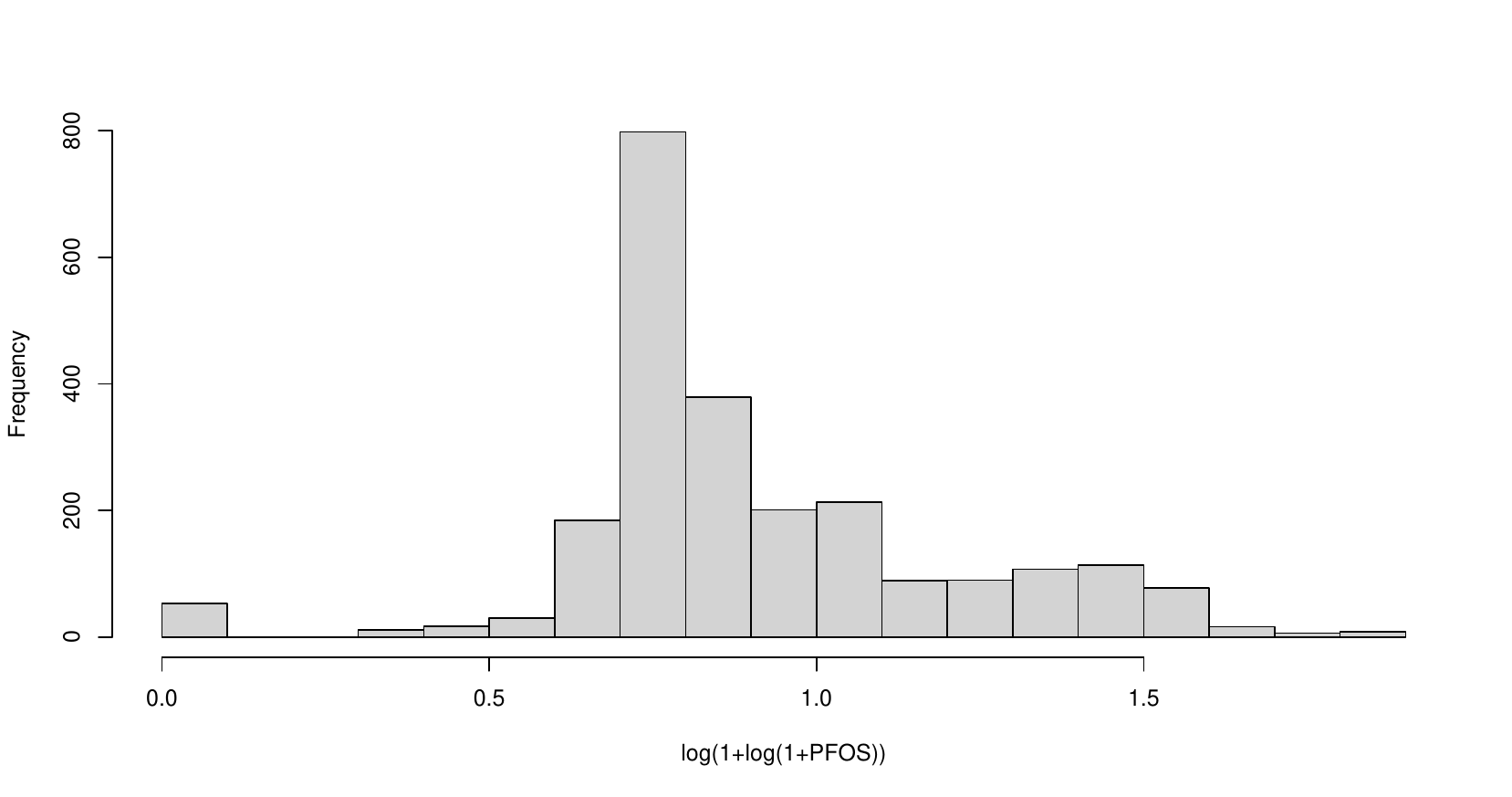}
    \includegraphics[width=0.49\linewidth]{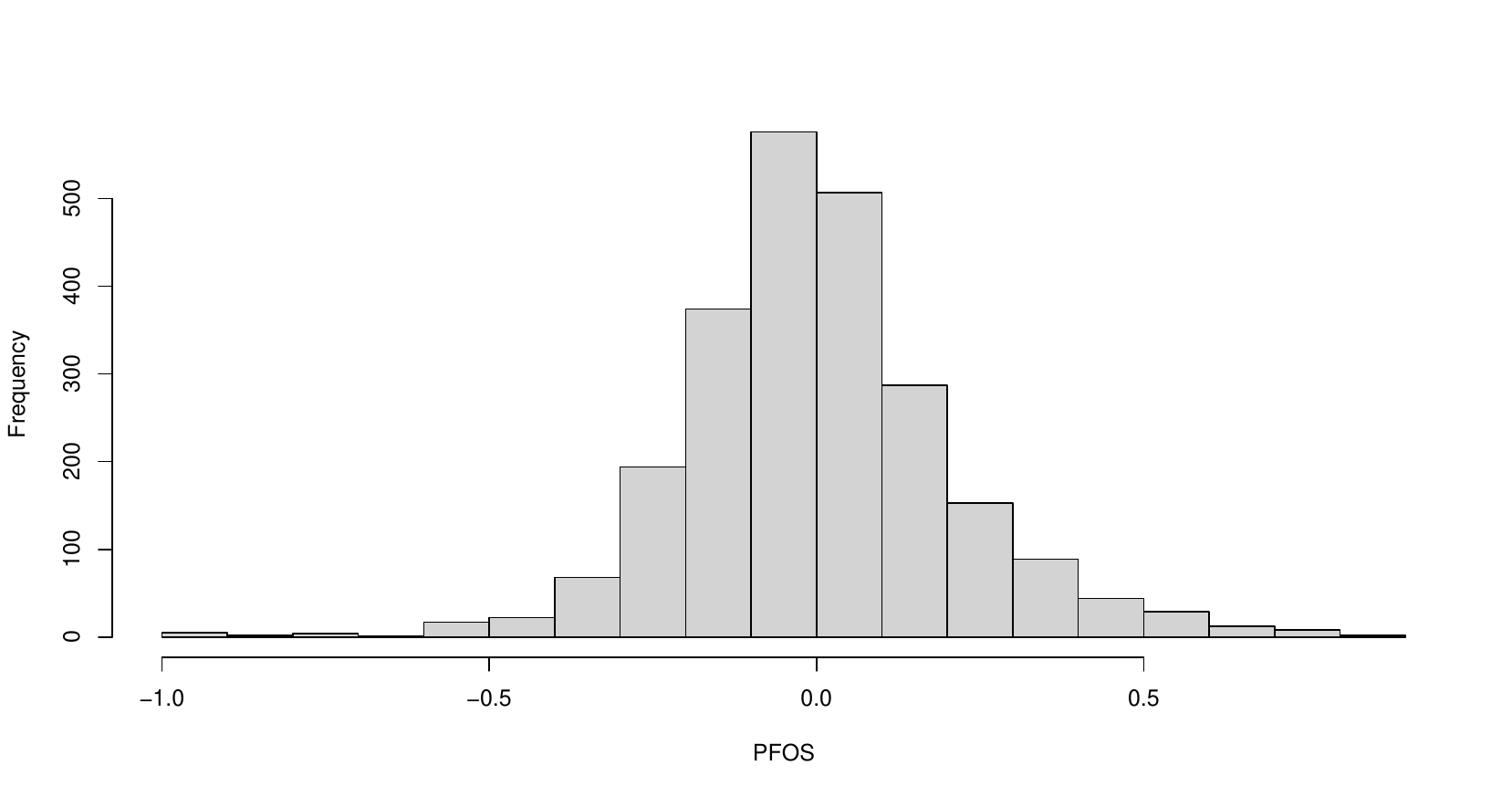}
    \caption{Histograms of PFOS data and residuals from simple linear regression (SLR) models across sites where the data are not censored. \textit{Top row}: Histogram of raw PFOS responses (left) and histogram of residuals from an SLR of raw PFOS on 268 selected covariates (right), both exhibiting strong positive skewness. \textit{Bottom row}: Histogram of iterated log-transformed PFOS responses, $g(\mbox{PFOS}) = \mbox{log}(1 + \mbox{log}(1 + \mbox{PFOS}))$ (left), and histogram of residuals from an SLR of the transformed response on the same set of covariates (right), which display an approximately Gaussian distribution.}
    \label{fig:eda_hist}
\end{figure}

The histogram of the raw PFOS responses, and that of the residuals from a simple linear regression (SLR) using the 261 selected covariates, are shown in the top row of Figure \ref{fig:eda_hist}. Both the raw data and the residuals exhibit strong positive skewness, violating Gaussian assumptions. To address this, we considered the iterated log-transformation, $g(\mbox{PFOS}) = \mbox{log}(1 + \mbox{log}(1 + \mbox{PFOS}))$, which provides a more suitable distributional fit. The histogram of the transformed PFOS data is displayed in the bottom-left panel of Figure \ref{fig:eda_hist}. Residuals from an SLR of the transformed response against the same set of covariates exhibit a Gaussian-like histogram (bottom-right panel of Figure \ref{fig:eda_hist}), supporting the appropriateness of the iterated log-transformation for subsequent modeling.

\begin{figure}
\centering
    \includegraphics[width=0.8\linewidth]{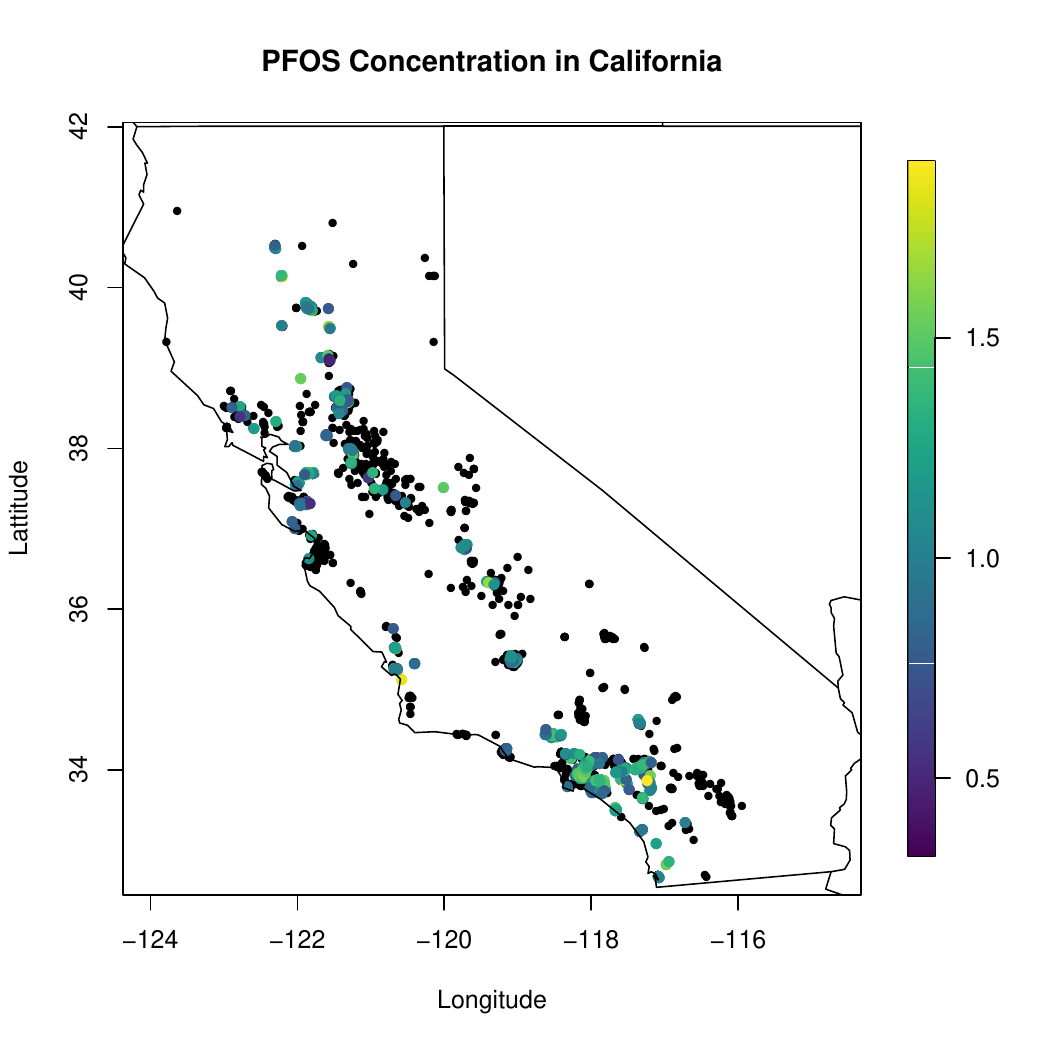}
    \caption{Concentrations of (transformed) PFOS, using the transformation $g(\textrm{PFOS}) = \log(1 + \log(1 + \textrm{PFOS}))$, measured at 2,394 irregularly-sampled spatial locations across the study area (in ng/L). The tiny black dots indicate the sites with censored data.} 
    \label{fig:all_locs}
\end{figure}

Figure \ref{fig:all_locs} shows the transformed PFOS concentration measurements after transforming the raw PFOS by $g(\textrm{PFOS}) = \log(1 + \log(1 + \textrm{PFOS}))$ at the 2,394 irregularly sampled spatial locations across the study domain. Since our study domain is unchanged from earlier work using this response variable \citep{sahoo2025computationally}, we use the same Mat\'{e}rn correlation structure with smoothness $\nu = 1$ given by \beq \label{e:matcor} C(\bs,\bs';\rho) \equiv C(d;\rho) = r\frac{d}{\rho}\kappa_1\Lp \frac{d}{\rho} \Rp + (1-r)\bbi (\bs = \bs'), \eeq to model the response. Here $d = ||\bs - \bs'||_2$ is the Euclidean distance between locations $\bm{s}$ and $\bm{s}', \rho > 0$ is the range parameter, $r \in [0, 1]$ is the ratio of spatial to total variation, $\kappa_1(\cdot)$ is the modified Bessel function of second kind with degree 1, and $\bbi(\cdot)$ is the indicator function. The computational mesh employed previously is also retained for the current analysis. Although these preliminary results are based only on the uncensored observations, they highlight the need for a full spatial model that explicitly accounts for censoring. In particular, a majority of measurements are censored, and disregarding them would bias predictions and degrade performance in neighboring areas.

\section{Statistical Methodology} \label{s:methods}

\subsection{Approximation Framework for Censored Spatial Data} \label{ss:meth_base}

Our modeling framework for high-dimensional censored spatial data builds on the approximation approach developed in \cite{sahoo2025computationally}. For completeness, we briefly summarize the main components before we extend the methodology here to incorporate variable selection.

Let $Y(\bs)$ denote the (transformed) PFOS concentration at a location $\bs \in \calD \subset \mathbb{R}^2$, with $p$ associated covariates $\bX(\bs) = (X_1(\bs),\ldots,X_p(\bs))^T$ and their corresponding effects represented by the coefficients $\bbeta_{(p \times 1)} = (\beta_1, \ldots, \beta_p)^T$. The spatial regression model takes the form
\begin{equation} \label{eq:base_model}
Y(\bs) = \bX(\bs)^T \bbeta + \sigma Z(\bs),
\end{equation}
where $Z(\bs)$ is a zero-mean Gaussian process with unit variance and correlation function $C(\bs,\bs';\rho)$, representing the spatial correlation incurred by the process between locations $\bs$ and $\bs'$. Here, $\sigma^2$ denotes the total variation present in the data. For $n$ observed sites $\calS = \lbrace \bs_1,\ldots,\bs_n \rbrace$, we can write the spatial regression model as
\begin{equation} \label{eq:matrix_model}
\bY = \bX\bbeta + \sigma \bZ, 
\end{equation}
where $\bY_{(n \times 1)}$ denotes the stacked vector of observations at the $n$ spatial locations, $\bX_{(n \times p)}$ is the matrix of covariates and $\bZ \sim \text{MVN}\big(\bzero, r\bSigma + (1 - r)\bI_n\big)$ with $\bSigma_{ij} = C(\bs_i,\bs_j;\rho)$, and $r \in (0,1)$ partitions the total variation into a spatially structured component and a nugget effect.

Now suppose that responses are censored below detection limits (left-censored) at a subset of sites $\calS^{(c)} = \lbrace \bs_1^{(c)}, \ldots, \bs_{n_c}^{(c)}\rbrace \subset \calS$ with corresponding censoring limits $\mathcal{U} = \lbrace u_1^c, \ldots, u_{n_c}^c \rbrace$,  while all responses are observed at the uncensored sites $\calS^{(obs)}$, such that $\calS = \calS^{(c)} \cup \calS^{(obs)}$. The censoring then results in the joint likelihood function for all the observations (censored and non-censored) to be
\begin{equation} \label{likelihood}
    \mathcal{L}(\bm{\theta}) = \int_{\bm{y}^{(c)} \leq \bm{u}} f_{\textrm{MVN}}(\bm{y}; \bm{X} \bm{\beta}, \sigma^2 [r \bm{\Sigma} + (1 - r) \bm{I}_{n}]) \hspace{0.05cm} d\bm{y}^{(c)},
\end{equation}
where the integral is over the censored responses $\by^{(c)} = \lbrace (\bs): \bs \in \calS^{(c)} \rbrace$, $f_{\textrm{MVN}}(\cdot; \bmu, \bSigma)$ denotes a multivariate normal density with mean $\bmu$ and covariance matrix $\bSigma$ and $\btheta = (\bbeta, \rho, \sigma^2, r)$. This likelihood is analytically intractable for large $n$, since it involves a high-dimensional integral of the multivariate Gaussian probability density over a truncated domain coupled with dense covariance structures. To simplify, we decompose the latent process into a structured spatial process with Mat\'ern correlation, $W(\bs)$, and a white noise process $\varepsilon(\bs)$. This gives us 
\begin{equation} \label{eq:split}
\bY = \bX\bbeta + \sigma \bW + \sigma \bm{\varepsilon},
\end{equation}
where $\bW \sim \text{MVN}(\bzero, r\bSigma)$ and $\bm{\varepsilon} \sim \text{MVN}(\bzero, (1-r)\bI_n)$. Under this formulation, the censored likelihood can be expressed as
\begin{eqnarray} \label{likelihood2}
   \nonumber  \mathcal{L}(\bm{\theta}) = \int \prod_{i: \bm{s}_i \notin \mathcal{S}^{(c)}}  (\sigma^2(1 - r))^{-1/2} \phi \left( \frac{y_i - \bm{x}_i^{\sf T}\bm{\beta} - w_i}{\sqrt{\sigma^2(1 - r)}} \right)\\
 \times \prod_{i: \bm{s}_i \in \mathcal{S}^{(c)}} \Phi\left( \frac{u_i - \bm{x}_i^{\sf T}\bm{\beta} - w_i}{\sqrt{\sigma^2(1 - r)}} \right) f_{\textrm{MVN}}(\bm{w}; \bm{0}, \sigma^{2} r \bm{\Sigma}) d\bm{w},
\end{eqnarray}
where $\Phi(\cdot)$ and $\phi(\cdot)$ are the standard normal distribution and density functions, respectively.

Even with this decomposition, evaluating the integral is computationally intensive for large $n$ due to dense covariance matrices. To circumvent this, we draw posterior samples from the joint density of $(\btheta, \bW)$ given by \beq \label{eq:jointdens}
\begin{split}
    \pi(\btheta,\bW) &=\prod_{i: \bm{s}_i \notin \mathcal{S}^{(c)}}  (\sigma^2(1 - r))^{-1/2} \phi \left( \frac{y_i - \bm{x}_i^{\sf T}\bm{\beta} - \bW_i}{\sqrt{\sigma^2(1 - r)}} \right)\\
 &\quad \times \prod_{i: \bm{s}_i \in \mathcal{S}^{(c)}} \Phi\left( \frac{u_i - \bm{x}_i^{\sf T}\bm{\beta} - \bW_i}{\sqrt{\sigma^2(1 - r)}} \right) f_{\textrm{MVN}}(\bm{W}; \bm{0}, \sigma^{2} r \bm{\Sigma}).\\
\end{split} \eeq This ensures that the posterior samples for both $\btheta$ and $\bW$ come from the corresponding marginal posterior distributions. However, evaluating the joint posterior is also cost prohibitive due to the large number of observations and cubic complexity associated with evaluating a multivariate normal density with dense covariance matrix. To address this, we follow the SPDE-based approximation of \cite{lindgren2011explicit}, as detailed in \cite{sahoo2025computationally}. In short, the latent Gaussian process $\bW$ is replaced by a Gaussian Markov random field $\tilde{\bW} = \bA \bW^*$ with $\bW^*$ having a sparse precision matrix $\bQ_\rho$. This sparsity enables efficient posterior computation while preserving the key spatial dependence structure. The resulting approximate spatial hierarchical model can be written as
\beq \label{e:finalmod} \begin{split}
    \bY|\bW^* &\sim \text{MVN}\Lp \bX\bbeta + \sigma r \bA\bW^*, \sigma^2(1-r)\bI_n \Rp\\
    \bW^* &\sim \text{MVN} (\bzero, \bQ_\rho^{-1})\\
    \LP \bbeta, \rho, \sigma^2,r \RP &\sim \pi(\bbeta) \times \pi(\rho) \times \pi(\sigma^2) \times \pi(r).
\end{split} \eeq

Thus, the approximate model retains the original censored spatial likelihood formulation while scaling to massive datasets by exploiting the sparsity of the SPDE–GMRF representation. In the following subsection, we extend this framework by developing and comparing multiple strategies for variable selection.

\subsection{Variable Selection Methodology}
\label{ss:meth_varsel}

When the number of candidate covariates is large, it is essential to identify which predictors are truly associated with the response, rather than allowing weak or spurious effects to inflate model complexity of the regression model. Within the approximate spatial hierarchical model introduced in \eqref{e:finalmod}, the covariate effects enter through the linear predictor $\bX\bbeta$. In this framework, estimation of regression effects is challenging because the censored likelihood provides limited information in regions of the response distribution below the detection threshold, and the spatial random effect captures structured residual variation that may overlap with covariate contributions. Consequently, regression coefficients may be weakly identified, leading to instability in posterior inference if left unregularized.

To address this, we employ shrinkage priors on 
$\bbeta$. Specifically, we adopt the Horseshoe+ prior of \cite{bhadra2017horseshoe+}, which is a global-local shrinkage prior. This enables adaptive regularization by allowing strong signals to remain relatively large while shrinking the effects of irrelevant or weak covariates toward zero. In particular, for $i = 1,\ldots,p$:
\begin{equation*}
    \begin{split}
        \beta_i | \lambda_i,\eta_i,\tau &\indep N(0,\lambda_i^2)\\
        \lambda_i | \eta_i, \tau &\indep C^+(0,\tau\eta_i)\\
        \eta_i | \tau &\iid C^+(0,1)\\
        \tau &\sim U(0,1),
    \end{split}
\end{equation*} where $C^+(0,q)$ denotes the half-Cauchy distribution with scale $q$. Integrating over the $\eta_i$, we get \[\pi(\lambda_i|\tau) = \frac{4}{\pi^2\tau} \frac{\log(|\lambda_i/\tau|)}{\Lp\lambda_i/\tau\Rp^2 - 1}.\] 

The prior specification adopted in \cite{sahoo2025computationally} assigns independent Gaussian priors to the regression coefficients $\beta_i$, which is equivalent to a ridge-type regularization and thus induces uniform shrinkage. While ridge regression is effective in controlling variance, it is well documented that such priors do not perform selective shrinkage; small coefficients are not sufficiently shrunk toward zero, while large and truly non-negligible effects are unduly attenuated. This lack of adaptivity motivates the use of the Horseshoe+ prior, which simultaneously allows strong shrinkage of near-zero coefficients while preserving large signals with minimal bias.

The remaining priors on $(\rho,\sigma^2,r)$ follow those in \cite{sahoo2025computationally}.  We provide the full list of prior choices below for completeness: 
\beqs \label{e:priors} \begin{split}
    \beta_i | \lambda_i,\eta_i,\tau &\indep N(0,\lambda_i^2)\\
        \lambda_i | \eta_i, \tau &\indep C^+(0,\tau\eta_i)\\
        \eta_i | \tau &\iid C^+(0,1)\\
        \tau &\sim U(0,1)\\
        \sigma^2 &\sim IG(0.1,0.1)\\
        \rho &\sim U\Lp 0,0.5\Delta_{\calS} \Rp\\
        r &\sim U(0,1),
\end{split} \eeqs with $\Delta_{\calS}$ being the maximum range of the spatial domain.

\subsubsection{Post-Processing Rules for Variable Inclusion} \label{sss:zerocoefmeth}

While shrinkage priors provide posterior distributions for the regression coefficients, they do not yield binary inclusion directly. To perform variable selection within the censored spatial framework, we therefore implement multiple post-processing strategies that translate posterior summaries into rules for covariate inclusion. These complementary approaches enable us to balance interpretability with statistical rigor, and allow us to assess the robustness of variable selection results under different decision rules, ensuring that retained covariates represent meaningful contributions to explaining variation in the response.

First, we consider a credible interval based method for variable selection, where a covariate is included if the 95\% posterior credible interval for the corresponding regression coefficient excludes zero. This rule is simple to interpret and directly reflects the posterior uncertainty of each coefficient, though it can be conservative in high-dimensional settings.

Next, we implement the idea of shrinkage weights thresholding, as suggested by \cite{bhadra2017horseshoe+}. We compute the shrinkage weights for coefficient $\beta_i$ as $\kappa_i = 1/(1+\lambda_i^2\tau^2)$, where $\lambda_i$ is a local scale parameter and $\tau$ is the global scale parameter. The value of $\kappa_i$ governs the effective amount of shrinkage applied to $\beta_i$. Specifically, when $\lambda_i^2\tau^2$ is small, $\kappa_i$ approaches one, implying strong shrinkage and pulling the corresponding coefficient toward zero. Conversely, when $\lambda_i^2\tau^2$ is large, $\kappa_i$ approaches zero, indicating weak shrinkage and allowing the coefficient to remain relatively large, as supported by the data. Thus, we include the $i$th variable ($i = 1,\ldots, p$) in the model if the effective shrinkage weight \[\mathbb{E}\Lp \kappa_i|\bY \Rp = \mathbb{E}\Lp 1/(1+\lambda_i^2\tau^2)|\bY \Rp < 1/2,\] where $\mathbb{E}(\cdot|\bY)$ represents the posterior mean. This criterion performs well for ultra-sparse scenarios, but can be conservative otherwise, as seen in the simulation studies in Section \ref{s:sim}.

Finally, we implement the sequential two-means clustering approach of \cite{li2017variable} to partition variables into significant and non-significant groups, utilizing the \texttt{VsusP} package \citep{chapagain2024variable} in \texttt{R}. For a given threshold parameter, this method classifies the absolute values of the coefficients, $|\beta_i|$s, into two clusters and continuously recalculates cluster means and readjusts cluster memberships
until a stable set of clusters are found. Unlike interval-based or shrinkage weight rules, this method explicitly treats selection as a clustering problem and adapts to moderate-sparsity settings.

\section{Simulation Studies}
\label{s:sim}

We conduct a comprehensive simulation study to evaluate the performance of the approximate spatial model proposed in Section \ref{ss:meth_base} and to compare the three variable selection strategies discussed in Section \ref{ss:meth_varsel}. Performance is assessed across $100$ replicate datasets for multiple distinct scenarios generated by varying the censoring level, spatial range parameter, signal-to-noise ratio (partial sill to sill ratio), and the proportion of effective covariates. 

The spatial domain is fixed to the unit square $[0,1]^2$ with observations placed on a $100 \times 100$ grid. We set the number of covariates to $p = 100$, which remains constant across all scenarios. We simulate data under $36$ different scenarios by considering (a) the censoring levels of $20\%$ and $45\%$, (b) range parameter values of $0.07$, $0.12$, and $0.20$, (c) partial sill to sill ratios of $0.91$ and $0.80$, and (d) proportions of non-significant covariates equal to $5\%$, $50\%$, and $95\%$ respectively. For each scenario, $100$ datasets are generated from the model in Eq. (\ref{eq:matrix_model}) with $\sigma^2 = 1$ and covariates $\bX$ drawn independently from a standard normal distribution. Observations are randomly censored according to the percentile cutoff corresponding to either the $20$th or $45$th percentile, matching the target censoring levels. The resulting censored datasets are then treated as observed data for model fitting.

We consider two aspects of performance of the model to be of primary interest: (i) predictive accuracy, and (ii) correct identification of significant covariates. To evaluate predictive accuracy, each dataset is randomly partitioned into a training set consisting of 80\% of the observations and a test set comprising the remaining 20\%. Within each training set, we consider the two different levels of censoring for the response by setting different values of the minimum detection limit. The spatial hierarchical model described in Section \ref{ss:meth_base} with priors from Section \ref{ss:meth_varsel} is then fit to the training set. Covariates are classified as significant or non-significant  based on the posterior samples of the corresponding $\beta$ parameters using each of the three methods in Section \ref{sss:zerocoefmeth}: (a) a credible interval–based rule (Cr), (b) shrinkage weights thresholding as proposed by \citet{bhadra2017horseshoe+} in conjunction with the horseshoe+ prior (HSP), and (c) the sequential two-means clustering rule implemented in the \texttt{R} package \texttt{VSusP} (2means). 

After identifying the significant covariates, predictions $\widehat{Y}(\bs)$ are computed for the test set, and the predictive performance for the $i$th dataset is summarized by the prediction root mean squared error
\[\text{prediction RMSE}_i = \sqrt{\sum_{\bs \in \text{Test Set}_i} (Y^*(\bs) - \hat{Y}(\bs))^2},\] where prediction RMSE$_i$ is the prediction RMSE for the $i$th dataset and $Y^*(\bs)$ denotes the true response value at location $\bs$. The overall prediction RMSE for a scenario is computed by averaging across datasets, while the standard deviation of ${\text{RMSE}_i}$ provides a measure of variability. Lower values of both metrics indicate better and more consistent predictive performance.

To assess variable selection accuracy, we measure the level of mismatch between the set of covariates identified as significant by each method (Cr, HSP, and 2-means) and the true generating set, using the Hamming distance as the evaluation metric. For covariate $k ~(k = 1, \ldots, 100)$ in dataset $i ~(i = 1, \ldots, 100)$ under method $j ~(j = 1, 2, 3)$, we define
$$
\gamma_{ijk} = \begin{cases} 0, \text{ if covariate } k \text{ is correctly classified as significant or non-significant,} \\
1, \text{ otherwise}.
\end{cases}
$$
The dataset-level mismatch percentage for method $j$ is then calculated as 
$$
\gamma_{ij} = 100 \times \frac{1}{100} \sum_{k=1}^{100} \gamma_{ijk},
$$
and the overall mismatch percentage is averaged across datasets as follows:
$$
\gamma_j = \frac{1}{100}\sum_{i=1}^{100} \gamma_{ij}.
$$
A smaller value of $\gamma_j$ indicates more accurate variable selection. As with prediction RMSE, we also report the standard deviation of ${\gamma_{ij}}$ across datasets as a measure of uncertainty in the mismatch percentages, with smaller values reflecting greater consistency.

\subsection{Simulation Results and Comparative Analysis}

Table \ref{tab:sim_RMSE_20} summarizes the predictive performance of the proposed method for 20\% censoring, combined with the three variable selection strategies (Cr, 2means, HSP) in terms of RMSE, evaluated across different scenarios defined by the range parameter, the signal-to-noise ratio (SNR), and the number of significant covariates. The corresponding results for the 45\% censoring case do not vary significantly and hence they are provided in Table \ref{tab:sim_RMSE_45} in \ref{app1}. Prediction RMSE values remain stable across censoring levels, generally around 2, 3, and 5 for the three different range parameters, regardless of SNR, number of significant covariates, or variable selection strategy. This stability indicates that censoring percentage has little effect on the predictive performance of the proposed method, which is consistent with the findings in \cite{sahoo2025computationally}.

\begin{table}
    \centering
    \caption{Table of prediction RMSEs for different combinations of range ($\rho$), partial sill to sill ratios (SNR), and percentage of zero $\beta$'s, holding the censoring level at $20\%$ for the proposed model coupled with three different variable selection strategies: credible interval based (Cr), two-means clustering based (2means), and horseshoe+ prior recommendation based (HSP). The estimate for prediction RMSE for each scenario is obtained by averaging over the corresponding $100$ datasets and the uncertainty (in brackets) is obtained by taking the standard deviation of the individual prediction RMSEs over the datasets. All numbers are rounded to two significant digits after the decimal.}
    \label{tab:sim_RMSE_20}
    \resizebox{1.05\textwidth}{!}{
    \begin{tabular}{cccccccc}
    \hline
    $\%$ of&Range& \multicolumn{3}{c}{SNR=$91\%$} & \multicolumn{3}{c}{SNR=$80\%$}\\
    \cline{3-5}    \cline{6-8}
    Zeros&$(\rho)$& Cr & 2means & HSP & Cr & 2means & HSP\\
    \hline
    \multirow{3}{*}{$95\%$} & $0.07$ &1.83 (0.07) &1.83 (0.07) &1.93 (0.07) &1.82 (0.07) &1.88 (0.07) &1.98 (0.06) \\
    & $0.12$ &3.02 (0.12) &3.02 (0.12) &3.08 (0.12) &3.05 (0.12) &3.05 (0.13) &3.11 (0.12) \\
    & $0.2$ &4.99 (0.19) &4.99 (0.19) &5.02 (0.18) &4.99 (0.19) &4.99 (0.19) &5.03 (0.19) \\
    \hline
    \multirow{3}{*}{$50\%$} & $0.07$ &1.83 (0.07) &1.83 (0.07) &5.16 (0.16) &1.88 (0.06) &1.88 (0.06) &5.16 (0.17) \\
    & $0.12$ &3.05 (0.10) &3.06 (0.10) &5.59 (0.17) &3.08 (0.11) &3.09 (0.11) &5.62 (0.17) \\
    & $0.2$ &5.02 (0.20) &5.02 (0.20) &6.86 (0.19) &4.99 (0.17) &5.00 (0.17) &6.84 (0.18) \\
    \hline
    \multirow{3}{*}{$5\%$} & $0.07$ &1.84 (0.06) &2.12 (0.11) &4.96 (0.22) &1.89 (0.07) &2.17 (0.11) &4.99 (0.20) \\
    & $0.12$ &3.04 (0.11) &3.23 (0.12) &5.52 (0.18) &3.10 (0.11) &3.27 (0.12) &5.60 (0.22) \\
    & $0.2$ &5.05 (0.19) &5.12 (0.20) &6.85 (0.24) &5.06 (0.21) &5.16 (0.22) &6.89 (0.28) \\
    \hline
    \end{tabular}
    }
\end{table}

Table \ref{tab:sim_HD_20} presents the performance of the method for 20\% censoring under the three variable selection strategies in terms of mismatch percentage, again across different combinations of range parameters, SNR levels, and numbers of significant covariates. The analogous results for 45\% censoring once again show similar qualitative patterns, and hence are relegated to Table \ref{tab:sim_HD_45} in \ref{app1}. In contrast to prediction RMSE, mismatch percentages are more sensitive to the choice of variable selection strategy, as well as to SNR, number of significant covariates, and range parameter.

\begin{table}
    \centering
    \caption{Table of mismatch percentages for different combinations of range ($\rho$), partial sill to sill ratios (SNR), and percentage of zero $\beta$'s, holding the censoring level at $20\%$ for the proposed model coupled with three different variable selection strategies: credible interval based (Cr), two-means clustering based (2means), and horseshoe+ prior recommendation based (HSP). The estimate for mismatch percentage for each scenario is obtained by averaging over the corresponding $100$ datasets and the uncertainty (in brackets) is obtained by taking the standard deviation of the individual mismatch percentages over the datasets. All numbers are rounded to two significant digits after the decimal.}
    \label{tab:sim_HD_20}
    \resizebox{1.05\textwidth}{!}{
    \begin{tabular}{cccccccc}
    \hline
    $\%$ of&Range& \multicolumn{3}{c}{SNR=$91\%$} & \multicolumn{3}{c}{SNR=$80\%$}\\
    \cline{3-5}    \cline{6-8}
    Zeros&$(\rho)$& Cr & 2means & HSP & Cr & 2means & HSP\\
    \hline
    \multirow{3}{*}{$95\%$} & $0.07$ &2.49 (1.23) &12.99 (12.35) &2.00 ($<0.01$) &2.44 (1.14) &11.41 (12.14) &2.00 ($<0.01$) \\
    & $0.12$ &2.02 (0.99) &5.60 (9.59) &2.00 ($<0.01$) &2.13 (1.12) &4.38 (8.45) &2.00 ($<0.01$) \\
    & $0.2$ &1.85 (0.98) &4.04 (7.15) &2.00 ($<0.01$) &2.23 (1.01) &2.45 (5.14) &2.00 ($<0.01$) \\
    \hline
    \multirow{3}{*}{$50\%$} & $0.07$ &1.80 (0.89) &1.28 (0.51) &28.71 (0.48) &1.62 (0.94) &1.22 (0.44) &28.63 (0.49) \\
    & $0.12$ &1.73 (0.97) &1.53 (0.61) &28.31 (0.47) &1.59 (0.79) &1.55 (0.58) &28.36 (0.48) \\
    & $0.2$ &1.95 (1.01) &2.34 (0.84) &28.33 (0.54) &1.87 (1.00) &2.21 (0.82) &28.30 (0.56) \\
    \hline
    \multirow{3}{*}{$5\%$} & $0.07$ &2.38 (0.72) &14.30 (2.05) &34.80 (0.84) &2.41 (0.70) &14.58 (2.18) &34.86 (0.84) \\
    & $0.12$ &3.25 (0.70) &14.66 (2.21) &34.81 (0.75) &3.51 (0.81) &14.31 (2.47) &35.02 (0.91) \\
    & $0.2$ &4.99 (0.99) &13.54 (3.11) &34.97 (0.98) &4.93 (0.95) &14.15 (2.73) &35.11 (1.12) \\
    \hline
    \end{tabular}
    }
\end{table}

\subsection{Key Takeaways from Simulation Studies}

The prediction RMSEs remain largely unaffected by censoring levels, the number of significant covariates, or different values of the partial sill to sill ratio (SNR), as shown in Tables \ref{tab:sim_RMSE_20} and \ref{tab:sim_RMSE_45}. However, they do exhibit a clear worsening trend with increasing range across all models. This pattern is typical in spatial analysis when data are split into training and testing sets, since partitioning can disrupt spatial signals and weaken predictive accuracy. Importantly, the scale of the prediction RMSE is relatively small compared to the variability of the actual observations, which have an interquartile range of about 10 – 12, with higher variability at larger range values. Across the variable selection strategies, the Cr and 2means methods yield fairly comparable prediction RMSE, while the HSP method tends to perform slightly worse. The standard deviations of the RMSE also display consistent behavior across settings, generally increasing with the range parameter while remaining stable with respect to censoring levels, SNR, and covariate sparsity. The key implication here is that both the Cr and 2means based variable selection methods achieve similar predictive accuracy, whereas the HSP-based method performs comparatively worse.

The mismatch percentages reveal a more nuanced set of patterns, as summarized in Tables \ref{tab:sim_HD_20} and \ref{tab:sim_HD_45}. Unlike RMSE, mismatches are not substantially influenced by censoring levels or SNR, but they are highly sensitive to the number of significant covariates, the choice of variable selection method, and the range parameter. For the Cr-based method, mismatch rates worsen as the range increases, while for the HSP-based method they remain largely stable regardless of range. The 2means-based method shows behavior that varies sharply with sparsity; in ultra-sparse settings, performance improves with increasing range, in moderate settings, mismatches worsen with range, and in dense settings, mismatches remain largely unchanged. Overall, the Cr-based method provides the best performance in terms of minimizing mismatches, particularly in moderate and dense settings. Exceptions arise in ultra-sparse cases, where the HSP-based method delivers outstandingly consistent results. However, the HSP-based method fails to maintain accuracy in moderate or dense cases, even though its results remain internally consistent. Moreover, the HSP-based method consistently underestimates the number of significant variables, indicating a systematic bias derived from the stringent inclusion criterion. The 2means-based method offers the best performance in the moderate sparsity setting but underperforms in ultra-sparse and dense scenarios. The key takeaway here is that the credible interval–based method is the most reliable overall, excelling particularly in dense settings. While the horseshoe+ prior–based method is best suited for ultra-sparse scenarios, it has a tendency to induce additional sparsity into the model which can bring about false negatives.

\section{Application to PFAS Data} \label{s:data_ana}

\subsection{Data Analysis Setup}

We applied the proposed modeling framework to the PFOS dataset with censoring, using the full set of 261 covariates. Variable selection was carried out using the three strategies described previously: the credible interval-based method (Cr), the two-means clustering method (2means), and the shrinkage weights thresholding associated with the horseshoe+ prior (HSP). As demonstrated in the simulation studies, the predictive power of these strategies does not vary substantially, and in the absence of a known ground truth in the real data setting, direct measures of mismatch cannot be computed. Therefore, rather than committing to a single strategy, we explore all three methods in parallel. Since this is an exploratory analysis, we report the full set of potential drivers of PFOS concentrations in California groundwater as identified by each method.

The model was implemented with default hyperparameter settings, specifically $\beta_i^{(0)} = 0$, $a_\sigma = b_\sigma = 0.1$, and $\Delta_{\text{max}} = 0.5\Delta_S$, where $\Delta_S$ is the maximum pairwise distance between observation sites. The responses were $\mbox{log}(1 + \mbox{log}(1 + \mbox{PFOS}))$ levels, and all computations and summaries were based on this transformation rather than the original PFOS levels. For the analysis, we ran a chain of 100,000 iterations with thinning set to 10, resulting in 10,000 posterior samples. The first 2,000 samples were then discarded as burn-in samples. Computations were performed on UNIX servers with 4 GB RAM, and the full analysis took approximately 8.5 hours to complete.


\subsection{Data Analysis Results \& Interpretation}

The credible interval based method (Cr) selected $23$ significant variables, the means and standard deviations for which are presented in Table \ref{tab:asda_beta_cr} (multiplied by 1000, for ease of reading). These include number of samples taken, socioeconomic variables such as median rent, percentage of people on medicaid, gender (female), race (American Indians, Other races), land cover type (herbs), industries present (metal coating, textile and leather, oil and gas, cement manufacturing), distance to several industries (oil and gas, textile and leather, industrial gas, cement manufacturing, electronics, chemical manufacturing), airports (part 139) and distance to airport, traffic density, and elevation of the well site.

The two-means based method (2means) selected $8$ significant variables and the intercept, the means and standard deviations for which are presented in Table \ref{tab:asda_beta_2means}. The selected variables include socioeconomic variable (local GINI index), race (American Indian), change in land cover types (barren and woody wetland types), number of industries (cement manufacturing industry), number of airports (part 139) near the site, and the ozone concentration around the site.

\begin{table}
    \centering
    \caption{Estimates (posterior means) and uncertainty measures (posterior standard deviations) for the variables selected to be associated with PFOS concentration in the study area within California by the credible interval based method (Cr). All values are multiplied by $1,000$ and rounded to 3 significant digits post decimal for ease of representation.}
    \label{tab:asda_beta_cr}
    \begin{tabular}{cccc}
    \hline
    Variable & Coefficient & Estimates & SD\\
    \hline
    Number of Samples & $\beta_{2}$& $14.132$& $1.784$\\
    Median rent within 1 km buffer &$\beta_{11}$& $0.331$& $0.129$\\
    $\%$ people using Medicaid within 5 km buffer &$\beta_{26}$& $24.549$& $9.031$\\
    Female ($\%$) within 1 km buffer &$\beta_{67}$& $18.199$& $6.631$\\
    Female ($\%$) within 5 km buffer &$\beta_{68}$& $-42.593$& $17.656$\\
    American Indian ($\%$) within 5 km buffer &$\beta_{72}$& $169.461$& $64.242$\\
    Other races ($\%$) within 1 km buffer &$\beta_{77}$& $-7.020$& $3.405$\\
    $\%$ land covered by herb within 1 km buffer &$\beta_{135}$& $-12.696$& $3.923$\\
    Number of metal coating factories in 1 km buffer &$\beta_{146}$& $-34.274$& $14.276$\\
    Number of metal coating factories in 5 km buffer& $\beta_{147}$& $6.848$& $3.282$\\
    Number of Textile and Leather factories in 1 km buffer&$\beta_{154}$& $-66.613$& $29.500$\\
    Number of oil and gas factories in 5 km buffer& $\beta_{171}$& $23.825$& $10.935$\\
     Number of cement manufacturing factories in 5 km buffer& $\beta_{177}$ & $115.998$&$55.407$ \\
     Number of Airports (Part 139) within 1 km buffer& $\beta_{188}$ &  $359.554$& $174.835$\\
    Distance to the nearest oil and gas factory & $\beta_{191}$& $-0.017$& $0.007$\\
    Distance to nearest textile and leather factory& $\beta_{194}$& $-0.030$& $0.009$\\
     Distance to nearest industrial gas factories& $\beta_{204}$ & $0.015$& $0.007$\\
    Distance to nearest electronics industry& $\beta_{207}$& $0.030$& $0.009$\\
    Distance to nearest chemical manufacturing industry&$\beta_{210}$& $0.030$& $0.011$\\
     Distance to nearest cement manufacturing factory& $\beta_{211}$ & $0.011$& $0.005$\\
    Distance to nearest airport& $\beta_{213}$& $0.022$& $0.009$\\
    Traffic density within 150 m of the census tract boundary&$\beta_{229}$& $-0.097$& $0.033$\\
    Elevation & $\beta_{255}$& $-1.891$& $0.756$\\
    \hline
    \end{tabular}
\end{table}

\begin{table}
    \centering
    \caption{Estimates (posterior means) and uncertainty measures (posterior standard deviations) for the variables selected to be associated with PFOS concentration in the study area within California by the sequential two-means clustering based method (2means). All values are rounded to 3 significant digits post decimal. Variables marked with the $^\dag$ symbol were also selected by the horseshoe$+$ prior based strategy after loosening the cutoff from $0.5$ to $0.9$.}
    \label{tab:asda_beta_2means}
    \begin{tabular}{cccc}
    \hline
    Variable & Coefficient & Estimates & SD\\
    \hline
    Intercept$^\dag$ & $\beta_{1}$& $0.435$& $1.512$\\
    GINI index within 5 km buffer$^\dag$ & $\beta_{8}$& $-0.412$& $1.152$\\
    American Indian ($\%$) within 5 km buffer &$\beta_{72}$& $0.169$& $0.064$\\
    \% change in barren landcover in 1 km buffer & $\beta_{111}$& $0.200$& $0.125$\\
    \% change in woody wetland cover in 1 km buffer$^\dag$ & $\beta_{115}$& $-0.630$& $1.149$\\
    \% change in woody wetland cover in 5 km buffer$^\dag$ & $\beta_{116}$& $-0.208$& $1.846$\\
    Number of cement manufacturing factories in 1 km buffer& $\beta_{176}$ & $-0.240$&$0.249$ \\
     Number of Airports (Part 139) within 1 km buffer& $\beta_{188}$ &  $0.360$& $0.175$\\
    Ozone concentration$^\dag$ & $\beta_{216}$& $0.623$& $5.195$\\
    \hline
    \end{tabular}
\end{table}

The behavior of the three selection strategies were very different from each other. The horseshoe$+$ prior based recommendation (HSP) method did not pick any covariates to be significant when the cutoff for the effective shrinkage weights was set at 0.5. As we have seen in the simulation studies (Section \ref{s:sim}), the HSP method has a tendency to underestimate number of significant variables systematically due to perhaps a strict inclusion criterion. Therefore, not selecting any covariates as significant is not at all unexpected behavior from this selection strategy. Upon loosening the inclusion criteria to have the effective shrinkage weights smaller than $0.9$ instead of the recommended $0.5$, it selected $5$ variables to be significant, all of which were also selected by the `2means' method (marked with a $^\dag$ symbol in Table \ref{tab:asda_beta_2means}).

The variables selected by the two-means based method (2means) have one commonality, their magnitude is big. This is in line with the selection procedure for the `2means' method since it uses a two-means clustering of variables based on the magnitude of coefficients, resulting in coefficients with large magnitudes getting selected. However, the 2means method does not seem to incorporate the posterior standard deviation of these coefficients into the selection strategy well as several of these selected variables have high standard errors and would be deemed insignificant in the classical sense.

The variables selected by the credible interval based method (Cr) on the other hand is guaranteed to be statistically significant in the classical sense. This leads to quite a few variables being selected. Since the horsehsoe$+$ prior has already provided some shrinkage to all the coefficients, the increased number of covariates selected is probably not inaccurate, however, some of them do have a very small magnitude (estimates needed to be multiplied by 1000 before reporting).

Both variable selection methods (Cr and 2means) identified a set of related demographic, socioeconomic, industrial, environmental, and geographic variables to be significantly associated with PFOS concentrations in California groundwater. For instance, sites with a greater number of samples taken tended to show higher PFOS levels, which likely reflects targeted resampling in suspected hotspots rather than a direct causal effect. Socioeconomic factors such as higher local median rent and a greater proportion of residents enrolled in Medicaid within a 5 km buffer were positively associated with PFOS concentrations. In contrast, the GINI index within 5 km displayed a negative association, indicating that greater income inequality was linked to lower PFOS concentrations. These associations may proxy for patterns of urbanization, infrastructure age, and historical inequities in exposure, which are consistent with the environmental justice dimensions of PFAS contamination. Demographic composition also emerged as important: areas with higher percentages of American Indian populations within 5 km of sited exhibited higher PFOS levels, while communities with greater proportions of `Other' races showed lower levels. Such demographic indicators can be interpreted as markers of exposure context, reflecting historical land use and groundwater reliance, rather than biological susceptibility. In addition, the percentage of females within 1 km was positively associated with PFOS concentrations, whereas the percentage within 5 km showed a negative association. This contrasting pattern across spatial buffers may reflect underlying heterogeneity within settlement structures, suggesting that these variables act primarily as proxies for community composition rather than biological determinants.

Land cover patterns were also found to be significant. A higher percentage of herbaceous cover within 1 km was associated with lower PFOS, suggesting that vegetated or less impervious landscapes may mitigate contaminant transport into groundwater. Similarly, changes in barren land cover correlated with higher PFOS, while gains in woody wetlands were associated with lower levels, highlighting the role of vegetation in modulating PFAS mobility. Topography further influenced contamination patterns, with lower elevations (valley bottoms and depositional basins) exhibiting higher PFOS, consistent with groundwater flow and accumulation along alluvial aquifers in California’s Central Valley.

Finally, industrial activities were identified to be central determinants of PFOS levels. Proximity to oil and gas facilities, and textile and leather facilities were strongly associated with higher concentrations, with sites closer to such facilities exhibiting greater contamination. Counts of textile and leather factories, as well as metal coating and cement manufacturing industries, were also influential, though with nuanced spatial patterns; nearby counts sometimes had negative associations, while broader-area counts were positive. Airports, and particularly Part 139 commercial airports, emerged as especially important predictors: the number of airports within a 1 km buffer was strongly positively associated with PFOS levels, reinforcing the long-established link between aqueous film forming foam (AFFF) use and PFAS contamination in groundwater. Traffic density appeared less consistent as a predictor, with lower PFOS concentrations observed in sites immediately adjacent to very dense road corridors, suggesting that sites immediately adjacent to very dense traffic may not be the primary PFOS signal in this groundwater dataset.

The results of this study are consistent with findings from prior research on PFAS contamination. Numerous studies have established airports and military bases as key sources of PFAS  due to historical AFFF use, with groundwater plumes dominated by PFOS/PFHxS and precursors documented at and downstream of fire‑training areas and airfields \citep{hu2016detection, anderson2018occurrence, sunderland2019review}. Industrial sources such as textile and leather production, metal plating, and electronics manufacturing are also well-documented PFAS emitters \citep{kotthoff2015perfluoroalkyl, gluge2020overview}. The environmental drivers identified by our model match known transport processes. Sites with more herbaceous cover tend to have lower PFOS, likely because organic matter binds the chemical and slows its movement. In contrast, lower‑elevation alluvial areas show higher concentrations, consistent with contaminant accumulation in zones where water and sediments converge \citep{higgins2006sorption, ahrens2015stockholm, anderson2018occurrence}. Finally, the observed associations with demographic indicators such as the proportion of American Indian residents, gender composition, and Medicaid enrollment, are consistent with documented inequities in PFAS exposure among marginalized communities, who are reliant on groundwater as their primary source of drinking water \citep{sunderland2019review, smalling2023per}. These patterns also align with results obtained from analysis of NHANES data, which likewise reveal systematic variation in PFAS concentrations by sex and race/ethnicity \citep{kato2014changes}. These patterns parallel the associations observed in our model, in which demographic covariates may function as proxies for underlying exposure contexts such as differences in water source, occupational patterns, or residential proximity to contamination sources.

\section{Conclusions} \label{s:conc}
In this paper, we develop a Bayesian hierarchical framework to address the joint challenges posed by spatially dependent, left-censored responses, and a high-dimensional, heterogeneous set of candidate predictors. Building on a scalable SPDE–GMRF approximation, we incorporate global–local shrinkage priors that facilitate adaptive regularization and systematic variable selection. For variable selection, we evaluate three complementary strategies: a credible interval–based rule, a shrinkage weight thresholding approach, and a sequential two-means clustering method. Our simulation studies highlight the trade-offs among these strategies, with the credible interval rule emerging as the most reliable overall, the shrinkage weight method excelling in ultra-sparse settings, and the clustering approach offering flexibility in moderately sparse contexts. Applied to PFOS concentrations in California groundwater, the method identifies a parsimonious yet scientifically coherent set of demographic, industrial, and environmental predictors, offering both methodological advances for censored, spatial regression and insights into the factors influencing PFAS concentration.

While the horseshoe$+$ prior provides effective shrinkage and yields stable prediction RMSEs across all scenarios and variable selection strategies (Section \ref{s:sim}), the variable selection strategies exhibit notable differences in their identification of significant covariates. As seen in both simulations (Section \ref{s:sim}) and the real data application (Section \ref{s:data_ana}), the horseshoe$+$ prior has a systematic tendency to underestimate the number of significant covariates, reflecting its conservative inclusion criterion. The sequential two-means clustering approach, in turn, performs poorly in dense settings and is not optimal for ultra-sparse scenarios. Its reliance on coefficient magnitudes, without sufficiently accounting for posterior uncertainty, often leads to the selection of variables with large estimates but high variability (Section \ref{s:data_ana}). By contrast, the credible interval–based method demonstrate consistently strong performance across simulation regimes (Section \ref{s:sim}), and, although it identifies the largest number of significant covariates in the PFOS analysis, the prior-induced shrinkage from the horseshoe$+$ specification ensures that this outcome remains well-calibrated and scientifically interpretable.


This study is not without limitations, which we outline next. First, our framework fits a global model across all available locations, assuming that the same set of drivers govern PFOS concentrations statewide. However, in reality, sources of PFAS contamination may act more locally, and region-specific local models could provide insights into heterogeneous drivers. Second, our data are restricted to 40 counties in California with substantial reliance on groundwater, and thereby excluding major metropolitan areas. Thus, the results should be interpreted within this geographic scope, and future analyses incorporating a more comprehensive statewide dataset would be valuable. From a methodological standpoint, we have employed an isotropic Mat\'ern correlation structure with fixed smoothness. Extensions allowing anisotropy or spatially varying smoothness parameters could provide more flexibility and better capture directional dependence in PFAS transport. Finally, while our analysis focuses on PFOS, the dataset contains multiple PFAS compounds with varying levels of censoring, and a multivariate modeling framework that jointly accounts for their dependence structures could yield richer insights.

\appendix

\setcounter{table}{0}
\setcounter{figure}{0}

\section{Additional tables from the Simulation Study}
\label{app1}

We present the results for the performance of the proposed method with the three different variable selection strategies (Cr, 2means, HSP) across different range parameters, different partial sill to sill ratios and across different proportion of significant covariates when the censoring level is fixed at $45\%$. Table \ref{tab:sim_RMSE_45} presents prediction RMSEs (averaged over datasets) and Table \ref{tab:sim_HD_45} presents mismatch percentage (averaged over datasets), while the corresponding uncertainty quantification is presented in brackets by computing the standard deviation across datasets. The results do not lead to any different inference than those presented in Section \ref{s:sim}.

\begin{table}[h]
    \centering
    \caption{Table of prediction RMSEs for different combinations of range ($\rho$), partial sill to sill ratios (SNR), and percentage of zero $\beta$'s, holding the censoring level at $45\%$ for the proposed model coupled with three different variable selection strategies: credible interval based (Cr), two-means clustering based (2means), and horseshoe+ prior recommendation based (HSP). The estimate for prediction RMSE for each scenario is obtained by averaging over the corresponding $100$ datasets and the uncertainty (in brackets) is obtained by taking the standard deviation of the individual prediction RMSEs over the datasets. All numbers are rounded to two significant digits after the decimal.}
    \label{tab:sim_RMSE_45}
    \resizebox{1.05\textwidth}{!}{
    \begin{tabular}{cccccccc}
    \hline
    $\%$ of& Range& \multicolumn{3}{c}{SNR=$91\%$} & \multicolumn{3}{c}{SNR=$80\%$}\\
    \cline{3-5}    \cline{6-8}
     Zeros&$(\rho)$& Cr & 2means & HSP & Cr & 2means & HSP\\
    \hline
    \multirow{3}{*}{$95\%$} & $0.07$ &1.86 (0.06) &1.86 (0.06) &1.96 (0.06) &1.91 (0.07) &1.91 (0.07) &2.00 (0.07) \\
    & $0.12$ &3.08 (0.12) &3.08 (0.12) &3.15 (0.12) &3.11 (0.12) &3.11 (0.12) &3.17 (0.11) \\
    & $0.2$ &5.07 (0.19) &5.07 (0.19) &5.11 (0.19) &5.08 (0.19) &5.09 (0.19) &5.12 (0.19) \\
    \hline
    \multirow{3}{*}{$50\%$} & $0.07$ &1.88 (0.07) &1.88 (0.07) &5.10 (0.19) &1.93 (0.08) &1.94 (0.08) &5.12 (0.19) \\
    & $0.12$ &3.11 (0.13) &3.12 (0.12) &5.63 (0.18) &3.12 (0.12) &3.13 (0.13) &5.64 (0.18) \\
    & $0.2$ &5.12 (0.21) &5.13 (0.20) &6.92 (0.22) &5.11 (0.16) &5.12 (0.16) &6.94 (0.19) \\
    \hline
    \multirow{3}{*}{$5\%$} & $0.07$ &1.89 (0.07) &2.17 (0.11) &4.99 (0.21) &1.95 (0.07) &2.23 (0.11) &5.05 (0.21) \\
    & $0.12$ &3.13 (0.12) &3.29 (0.14) &5.60 (0.22) &3.19 (0.13) &3.37 (0.14) &5.65 (0.23) \\
    & $0.2$ &5.20 (0.19) &5.29 (0.19) &7.04 (0.25) &5.23 (0.18) &5.32 (0.19) &7.09 (0.26) \\
    \hline
    \end{tabular}
    }
\end{table}

\begin{table}
    \centering
    \caption{Table of mismatch percentages for different combinations of range ($\rho$), partial sill to sill ratios (SNR), and percentage of zero $\beta$'s, holding the censoring level at $45\%$ for the proposed model coupled with three different variable selection strategies: credible interval based (Cr), two-means clustering based (2means), and horseshoe+ prior recommendation based (HSP). The estimate for mismatch percentage for each scenario is obtained by averaging over the corresponding $100$ datasets and the uncertainty (in brackets) is obtained by taking the standard deviation of the individual mismatch percentages over the datasets. All numbers are rounded to two significant digits after the decimal.}
    \label{tab:sim_HD_45}
    \resizebox{1.05\textwidth}{!}{
    \begin{tabular}{cccccccc}
    \hline
    $\%$ of&Range& \multicolumn{3}{c}{SNR=$91\%$} & \multicolumn{3}{c}{SNR=$80\%$}\\
    \cline{3-5}    \cline{6-8}
    Zeros&$(\rho)$& Cr & 2means & HSP & Cr & 2means & HSP\\
    \hline
    \multirow{3}{*}{$95\%$} & $0.07$ &2.40 (1.06) &11.71 (12.19) &2.00 ($<0.01$) &2.19 (1.08) &12.10 (12.40) &2.00 ($<0.01$) \\
    & $0.12$ &2.43 (1.20) &2.67 (6.12) &2.00 ($<0.01$) &2.15 (1.18) &2.67 (6.13) &2.00 ($<0.01$) \\
    & $0.2$ &2.02 (0.96) &5.62 (6.85) &2.00 ($<0.01$) &1.85 (0.97) &6.88 (7.16) &2.00 ($<0.01$) \\
    \hline
    \multirow{3}{*}{$50\%$} & $0.07$ &1.73 (0.91) &1.31 (0.51) &28.49 (0.50) &1.57 (0.79) &1.37 (0.49) &28.48 (0.54) \\
    & $0.12$ &1.59 (0.89) &1.79 (0.67) &28.31 (0.51) &1.41 (0.70) &1.80 (0.72) &28.38 (0.49) \\
    & $0.2$ &2.02 (0.99) &2.77 (0.85) &28.22 (0.52) &1.94 (1.02) &2.44 (0.87) &28.32 (0.65) \\
    \hline
    \multirow{3}{*}{$5\%$} & $0.07$ &2.74 (0.66) &14.39 (2.54) &34.87 (0.84) &2.82 (0.70) &14.61 (2.20) &35.03 (0.90) \\
    & $0.12$ &3.89 (0.95) &13.79 (2.99) &34.94 (0.93) &3.95 (0.97) &14.53 (2.64) &35.02 (1.05) \\
    & $0.2$ &5.87 (1.07) &13.90 (2.72) &35.43 (1.08) &5.91 (0.94) &13.77 (3.22) &35.54 (1.22) \\
    \hline
    \end{tabular}
    }
\end{table}

\section{Descriptive Tables for the Covariates}
\label{app2}

We present two tables in this section. Table \ref{tab:app_industries} provides the number of identified industrial sectors potentially connected to per- and polyfluoroalkyl substances (PFAS) in California and within the study area that are included in the analysis. Table \ref{tab:app_covdes} provides a basic description of all covariates used in the analysis with the corresponding sources for them.

\begin{table}
\caption{Number of industries potentially related to per- and polyfluoroalkyl substances (PFAS) in the state of California and the study region.}
    \label{tab:app_industries}
    \centering
    \begin{tabular}{ccc}
    \hline
    Industry Sector & In California & In Study Area\\
    \hline
    Waste Management & 2711 & 1183\\
    Oil and gas & 624 & 268\\
    Metal coating & 1724 & 253\\
    Plastics and resins & 954 & 233\\
    Chemical manufacturing & 988 & 213\\
    Airports & 350 & 185\\
    Electronics industry & 2477 & 148\\
    Metal machinery manufacturing & 716 & 138\\
    Petroleum & 470 & 122\\
    Printing & 829 & 107\\
    Paints and coatings & 463 & 93\\
    Textiles and leather & 506 & 83\\
    Mining and refining & 112 & 58\\
    Cleaning product manufacturing & 316 & 54\\
    Furniture and carpet & 212 & 48\\
    Paper mills and products & 143 & 41\\
    National defense & 175 & 39\\
    Consumer products & 98 & 30\\
    Fire training & 61 & 30\\
    Cement manufacturing & 55 & 28\\
    Industrial gas & 132 & 25\\
    Glass products & 133 & 17\\
    Airports (part 139) & 31 & 15\\
    \hline
    Total & 14280 & 3411\\
    \hline
    \end{tabular}
\end{table}

\begin{longtable}{p{.2\textwidth} p{.55\textwidth} p{.12\textwidth} p{.12\textwidth}}
KILL&THIS&LINE&NOW \kill
\caption{Descriptive statistics (Sample mean and standard deviation) for the covariates present in the data and their sources}\\
\hline
Source & Name of Covariate & Mean & SD \\ 
  \hline
  \endfirsthead
  \caption[]{(Continued)}\\
  \hline
Source & Name of Covariate & Mean & SD \\ 
  \hline
  \endhead
GAMA & Number of samples tested at well for PFAS - PFOS & 4.93 & 5.84 \\ 
\hline
  US Census Bureau & median age within 1 km & 35.78 & 6.78 \\ 
   & median age within 5 km & 35.51 & 5.60 \\ 
  & median income (USD) within 1 km & 59462.40 & 22950.77 \\ 
  & median income (USD) within 5 km & 59592.04 & 18643.05 \\ 
  & GINI index of income inequality within 1 km & 0.42 & 0.05 \\ 
  & GINI index of income inequality within 5 km & 0.41 & 0.04 \\ 
  & average household size within 1 km & 3.17 & 0.60 \\ 
  & average household size within 5 km & 3.17 & 0.50 \\ 
  & median monthly rent (USD) within 1 km & 1252.75 & 367.65 \\ 
  & median monthly rent (USD) within 5 km & 1261.64 & 322.61 \\ 
  & median home value (USD) within 1 km & 328249.47 & 202293.68 \\ 
  & median home value (USD) within 5 km & 330712.41 & 183834.50 \\ 
  & \% population 85 or older within 1 km & 1.15 & 0.97 \\ 
  & \% population 85 or older within 5 km & 1.18 & 0.69 \\ 
  & \% population 65 or older within 1 km & 34.80 & 6.69 \\ 
  & \% population 65 or older within 5 km & 34.45 & 5.51 \\ 
  & \% population living with a disability within 1 km & 11.30 & 4.15 \\ 
  & \% population living with a disability within 5 km & 11.20 & 3.22 \\ 
  & \% households receiving SNAP/food stamp benefits within 1 km & 12.47 & 8.90 \\ 
  & \% households receiving SNAP/food stamp benefits within 5 km & 12.45 & 7.17 \\ 
  & \% population on public health insurance with medicare within 1 km & 24.17 & 10.61 \\ 
  & \% population on public health insurance with medicare within 5 km & 23.86 & 8.65 \\ 
  & \% population on public health insurance with medicaid within 1 km & 56.23 & 12.25 \\ 
  & \% population on public health insurance with medicaid within 5 km & 55.92 & 9.17 \\ 
  & \% population on public health insurance with VA health care within 1 km & 78.49 & 20.63 \\ 
  & \% population on public health insurance with VA health care within 5 km & 77.33 & 11.60 \\ 
  & \% population with a bachelor's degree or higher within 1 km & 23.53 & 14.53 \\ 
  & \% population with a bachelor's degree or higher within 5 km & 24.12 & 12.46 \\ 
  & \% population born outside of the US within 1 km & 25.02 & 13.09 \\ 
  & \% population born outside of the US within 5 km & 25.11 & 11.95 \\ 
  & unemployment rate within 1 km & 10.96 & 4.50 \\ 
  & unemployment rate within 5 km & 10.99 & 3.60 \\ 
  & \% population working in agriculture, forestry, fishing and hunting, or mining within 1 km & 4.72 & 8.64 \\ 
  & \% population working in agriculture, forestry, fishing and hunting, or mining within 5 km & 4.76 & 8.24 \\ 
  & \% population working in construction within 1 km & 6.73 & 2.84 \\ 
  & \% population working in construction within 5 km & 6.52 & 1.94 \\ 
  & \% population working in manufacturing within 1 km & 9.49 & 4.98 \\ 
  & \% population working in manufacturing within 5 km & 9.36 & 4.27 \\ 
  & \% population working in wholesale trade within 1 km & 3.21 & 1.82 \\ 
  & \% population working in wholesale trade within 5 km & 3.21 & 1.38 \\ 
  & \% population working in retail trade within 1 km & 11.49 & 3.11 \\ 
  & \% population working in retail trade within 5 km & 11.56 & 2.19 \\ 
  & \% population working in transportation and warehousing, and utilities within 1 km & 4.88 & 2.54 \\ 
  & \% population working in transportation and warehousing, and utilities within 5 km & 4.81 & 2.06 \\ 
  & \% population working in information within 1 km & 1.92 & 1.71 \\ 
  & \% population working in information within 5 km & 1.99 & 1.50 \\ 
  & \% population working in finance and insurance, and real estate and rental and leasing within 1 km & 5.20 & 2.52 \\ 
  & \% population working in finance and insurance, and real estate and rental and leasing within 5 km & 5.20 & 1.81 \\ 
  & \% population working in professional, scientific, and management, and administrative and waste management services within 1 km & 10.89 & 4.16 \\ 
  & \% population working in professional, scientific, and management, and administrative and waste management services within 5 km & 10.83 & 3.56 \\ 
  & \% population working in educational services, and health care and social assistance within 1 km & 20.39 & 5.76 \\ 
  & \% population working in educational services, and health care and social assistance within 5 km & 20.58 & 4.36 \\ 
  & \% population working in arts, entertainment, and recreation, and accommodation and food services within 1 km & 10.46 & 4.89 \\ 
  & \% population working in arts, entertainment, and recreation, and accommodation and food services within 5 km & 10.54 & 4.24 \\ 
  & \% population working in other services, except public administration within 1 km & 5.19 & 2.08 \\ 
  & \% population working in other services, except public administration within 5 km & 5.23 & 1.39 \\ 
  & \% population working in public administration within 1 km & 5.43 & 5.17 \\ 
  & \% population working in public administration within 5 km & 5.40 & 4.35 \\ 
  & \% population on private health insurance within 1 km & 56.78 & 16.82 \\ 
  & \% population on private health insurance within 5 km & 57.36 & 13.69 \\ 
  & \% population on public health insurance within 1 km & 35.86 & 10.60 \\ 
  & \% population on public health insurance within 5 km & 35.62 & 8.77 \\ 
  & \% population without health insurance within 1 km & 15.95 & 6.71 \\ 
  & \% population without health insurance within 5 km & 15.69 & 5.35 \\ 
  & \% population male within 1 km & 50.03 & 3.23 \\ 
  & \% population male within 5 km & 49.90 & 2.56 \\ 
  & \% population female within 1 km & 49.97 & 3.23 \\ 
  & \% population female within 5 km & 50.10 & 2.56 \\ 
  & \% population white within 1 km & 65.11 & 17.08 \\ 
  & \% population white within 5 km & 64.76 & 15.44 \\ 
  & \% population black or african-american within 1 km & 4.62 & 5.23 \\ 
  & \% population black or african-american within 5 km & 4.79 & 4.62 \\ 
  & \% population american indian or alaska native within 1 km & 0.94 & 1.50 \\ 
  & \% population american indian or alaska native within 5 km & 0.94 & 1.19 \\ 
  & \% population asian within 1 km & 10.75 & 11.37 \\ 
  & \% population asian within 5 km & 11.18 & 10.52 \\ 
  & \% population native hawaiian or other pacific islander within 1 km & 0.38 & 0.77 \\ 
  & \% population native hawaiian or other pacific islander within 5 km & 0.37 & 0.46 \\ 
  & \% population some other race within 1 km & 13.74 & 11.30 \\ 
  & \% population some other race within 5 km & 13.51 & 9.79 \\ 
  & \% population reporting two or more races within 1 km & 4.46 & 2.47 \\ 
  & \% population reporting two or more races within 5 km & 4.45 & 1.88 \\ 
  & \% population of hispanic or latino origin within 1 km & 42.13 & 23.60 \\ 
  & \% population of hispanic or latino origin within 5 km & 41.79 & 21.05 \\ 
  & \% population living in owner-occupied unit within 1 km & 52.92 & 16.73 \\ 
  & \% population living in owner-occupied unit within 5 km & 52.53 & 11.70 \\ 
  & \% population living in renter-occupied unit within 1 km & 45.09 & 16.28 \\ 
  & \% population living in renter-occupied unit within 5 km & 45.27 & 11.25 \\ 
  & \% population veterans within 1 km & 4.84 & 2.84 \\ 
  & \% population veterans within 5 km & 4.79 & 2.54 \\ 
  & average home age in years (as of 2015) within 1 km & 39.84 & 12.72 \\ 
  & average home age in years (as of 2015) within 5 km & 40.31 & 10.22 \\ 
  & total population of buffer (as of 2015) within 1 km & 5259.83 & 4813.32 \\ 
  & total population of buffer (as of 2015) within 5 km & 117478.58 & 99796.31 \\
  \hline
  GeoMAC Wildfire Perimeter Database & \% buffer 1 km area burned during 2008 to 2017 & 1.26 & 5.61 \\ 
  & \% buffer 5 km area burned during 2008 to 2017 & 0.62 & 5.04 \\ 
  \hline
   National Land Cover Database & \% buffer 1 km covered by urban impervious surface & 40.75 & 25.05 \\ 
  & \% buffer 5 km covered by urban impervious surface & 34.55 & 22.70 \\ 
    & \% buffer 1 km with no data & 0.00 & 0.00 \\ 
    & \% buffer 5 km with no data & 0.00 & 0.00 \\ 
    & \% buffer 1 km which did not change & 92.64 & 11.17 \\ 
    & \% buffer 5 km which did not change & 92.32 & 7.88 \\ 
    & \% buffer 1 km with new water landcover & 0.22 & 1.14 \\ 
    & \% buffer 5 km with new water landcover & 0.29 & 0.94 \\ 
    & \% buffer 1 km with new urban landcover & 5.91 & 10.15 \\ 
    & \% buffer 5 km with new urban landcover & 5.15 & 5.79 \\ 
    & \% buffer 1 km with landcover change within wetland class & 0.03 & 0.26 \\ 
    & \% buffer 5 km with landcover change within wetland class & 0.03 & 0.17 \\ 
    & \% buffer 1 km with new herbaceous wetland landcover & 0.02 & 0.13 \\ 
    & \% buffer 5 km with new herbaceous wetland landcover & 0.02 & 0.06 \\ 
    & \% buffer 1 km with landcover change within agriculture class & 0.08 & 0.50 \\ 
    & \% buffer 5 km with landcover change within agriculture class & 0.12 & 0.39 \\ 
    & \% buffer 1 km with new cultivated crop landcover & 0.52 & 2.46 \\ 
    & \% buffer 5 km with new cultivated crop landcover & 0.67 & 1.89 \\ 
    & \% buffer 1 km with new hay/pasture landcover & 0.04 & 0.25 \\ 
    & \% buffer 5 km with new hay/pasture landcover & 0.03 & 0.08 \\ 
    & \% buffer 1 km with new rangeland herbaceous and shrub landcover & 0.28 & 2.52 \\ 
    & \% buffer 5 km with new rangeland herbaceous and shrub landcover & 0.84 & 3.56 \\ 
    & \% buffer 1 km with new barren landcover & 0.03 & 0.18 \\ 
    & \% buffer 5 km with new barren landcover & 0.04 & 0.28 \\ 
    & \% buffer 1 km with new forest landcover & 0.25 & 2.71 \\ 
    & \% buffer 5 km with new forest landcover & 0.50 & 2.98 \\ 
    & \% buffer 1 km with new woody wetlands landcover & 0.00 & 0.05 \\ 
    & \% buffer 5 km with new woody wetlands landcover & 0.00 & 0.01 \\ 
    & \% buffer 1 km with no data & 0.00 & 0.00 \\ 
    & \% buffer 5 km with no data & 0.00 & 0.00 \\ 
    & \% buffer 1 km with open water land cover class & 0.64 & 2.71 \\ 
    & \% buffer 5 km with open water land cover class & 1.35 & 4.17 \\ 
    & \% buffer 1 km with perennial ice/snow land cover class & 0.00 & 0.00 \\ 
    & \% buffer 5 km with perennial ice/snow land cover class & 0.00 & 0.00 \\ 
    & \% buffer 1 km with developed, open space land cover class & 7.96 & 7.04 \\ 
    & \% buffer 5 km with developed, open space land cover class & 7.26 & 4.19 \\ 
    & \% buffer 1 km with developed, low intensity land cover class & 13.49 & 9.41 \\ 
    & \% buffer 5 km with developed, low intensity land cover class & 12.23 & 6.98 \\ 
    & \% buffer 1 km with developed, medium intensity land cover class & 34.74 & 23.09 \\ 
    & \% buffer 5 km with developed, medium intensity land cover class & 29.88 & 19.67 \\ 
    & \% buffer 1 km with developed, high intensity land cover class & 14.38 & 15.80 \\ 
    & \% buffer 5 km with developed, high intensity land cover class & 11.60 & 10.58 \\ 
    & \% buffer 1 km with barren land cover class & 0.77 & 3.04 \\ 
    & \% buffer 5 km with barren land cover class & 0.81 & 2.96 \\ 
    & \% buffer 1 km with deciduous forest land cover class & 0.05 & 0.41 \\ 
    & \% buffer 5 km with deciduous forest land cover class & 0.07 & 0.45 \\ 
    & \% buffer 1 km with evergreen forest land cover class & 1.90 & 9.72 \\ 
    & \% buffer 5 km with evergreen forest land cover class & 2.52 & 10.46 \\ 
    & \% buffer 1 km with mixed forest land cover class & 0.39 & 2.07 \\ 
    & \% buffer 5 km with mixed forest land cover class & 0.91 & 2.62 \\ 
    & \% buffer 1 km with shrub/scrub land cover class & 6.64 & 16.73 \\ 
    & \% buffer 5 km with shrub/scrub land cover class & 9.27 & 17.17 \\ 
    & \% buffer 1 km with grassland/herbaceous land cover class & 6.73 & 14.98 \\ 
    & \% buffer 5 km with grassland/herbaceous land cover class & 8.42 & 14.11 \\ 
    & \% buffer 1 km with pasture/hay land cover class & 1.46 & 5.91 \\ 
    & \% buffer 5 km with pasture/hay land cover class & 1.43 & 4.21 \\ 
    & \% buffer 1 km with cultivated crops land cover class & 9.49 & 21.32 \\ 
    & \% buffer 5 km with cultivated crops land cover class & 13.20 & 22.59 \\ 
    & \% buffer 1 km with woody wetlands land cover class & 0.52 & 2.02 \\ 
    & \% buffer 5 km with woody wetlands land cover class & 0.35 & 0.65 \\ 
    & \% buffer 1 km with emergent herbaceous wetlands land cover class & 0.83 & 3.31 \\ 
    & \% buffer 5 km with emergent herbaceous wetlands land cover class & 0.70 & 1.82 \\ 
    \hline
  EPA's PFAS Analytics Tool & number of all potential industry sources within buffer 1 km & 2.68 & 6.04 \\ 
    & number of all potential industry sources within buffer 5 km & 50.69 & 68.59 \\ 
    & number of electronics industry sites within buffer 1 km & 0.48 & 1.74 \\ 
    & number of electronics industry sites within buffer 5 km & 8.69 & 24.46 \\ 
    & number of metal coating sites within buffer 1 km & 0.51 & 1.70 \\ 
    & number of metal coating sites within buffer 5 km & 9.45 & 15.22 \\ 
    & number of mining and refining sites within buffer 1 km & 0.01 & 0.09 \\ 
    & number of mining and refining sites within buffer 5 km & 0.25 & 0.62 \\ 
    & number of printing sites within buffer 1 km & 0.21 & 0.74 \\ 
    & number of printing sites within buffer 5 km & 3.86 & 6.37 \\ 
    & number of plastic and resin sites within buffer 1 km & 0.21 & 0.74 \\ 
    & number of plastic and resin sites within buffer 5 km & 4.26 & 7.00 \\ 
    & number of textiles and leather sites within buffer 1 km & 0.14 & 0.55 \\ 
    & number of textiles and leather sites within buffer 5 km & 2.37 & 4.24 \\ 
    & number of chemical manufacturing sites within buffer 1 km & 0.15 & 0.52 \\ 
    & number of chemical manufacturing sites within buffer 5 km & 3.34 & 5.78 \\ 
    & number of paints and coatings sites within buffer 1 km & 0.13 & 0.45 \\ 
    & number of paints and coatings sites within buffer 5 km & 2.20 & 3.23 \\ 
    & number of petroleum sites within buffer 1 km & 0.11 & 0.41 \\ 
    & number of petroleum sites within buffer 5 km & 1.88 & 3.11 \\ 
    & number of waste management sites within buffer 1 km & 0.19 & 0.60 \\ 
    & number of waste management sites within buffer 5 km & 3.77 & 5.76 \\ 
    & number of cleaning product manufacturing sites within buffer 1 km & 0.06 & 0.29 \\ 
    & number of cleaning product manufacturing sites within buffer 5 km & 1.59 & 3.02 \\ 
    & number of paper mills and products sites within buffer 1 km & 0.04 & 0.23 \\ 
    & number of paper mills and products sites within buffer 5 km & 0.64 & 1.37 \\ 
    & number of industrial gas sites within buffer 1 km & 0.02 & 0.15 \\ 
    & number of industrial gas sites within buffer 5 km & 0.47 & 0.95 \\ 
    & number of oil and gas sites within buffer 1 km & 0.05 & 0.34 \\ 
    & number of oil and gas sites within buffer 5 km & 1.31 & 3.23 \\ 
    & number of airports within buffer 1 km & 0.05 & 0.34 \\ 
    & number of airports within buffer 5 km & 1.17 & 2.01 \\ 
    & number of national defense sites within buffer 1 km & 0.02 & 0.14 \\ 
    & number of national defense sites within buffer 5 km & 0.34 & 0.86 \\ 
    & number of cement manufacturing sites within buffer 1 km & 0.00 & 0.05 \\ 
    & number of cement manufacturing sites within buffer 5 km & 0.15 & 0.41 \\ 
    & number of furniture and carpet sites within buffer 1 km & 0.05 & 0.32 \\ 
    & number of furniture and carpet sites within buffer 5 km & 0.75 & 1.32 \\ 
    & number of metal machinery manufacturing sites within buffer 1 km & 0.18 & 0.62 \\ 
    & number of metal machinery manufacturing sites within buffer 5 km & 3.07 & 4.40 \\ 
    & number of glass products sites within buffer 1 km & 0.04 & 0.27 \\ 
    & number of glass products sites within buffer 5 km & 0.41 & 0.95 \\ 
    & number of fire training sites within buffer 1 km & 0.01 & 0.12 \\ 
    & number of fire training sites within buffer 5 km & 0.17 & 0.66 \\ 
    & number of consumer products sites within buffer 1 km & 0.01 & 0.12 \\ 
    & number of consumer products sites within buffer 5 km & 0.37 & 0.60 \\ 
    & number of airports (part 139) within buffer 1 km & 0.00 & 0.07 \\ 
    & number of airports (part 139) within buffer 5 km & 0.16 & 0.37 \\ 
    & distance to nearest potential PFAS source from any industry (m) & 2238.80 & 3547.55 \\ 
    & distance to nearest oil and gas site (m) & 16865.03 & 22296.39 \\ 
    & distance to nearest furniture and carpet site (m) & 14331.55 & 19191.51 \\ 
    & distance to nearest waste management site (m) & 4203.75 & 4274.98 \\ 
    & distance to nearest textiles and leather site (m) & 13555.09 & 18877.65 \\ 
    & distance to nearest printing site (m) & 14833.01 & 23195.76 \\ 
    & distance to nearest plastic and resin site (m) & 11740.97 & 17744.93 \\ 
    & distance to nearest petroleum site (m) & 10440.03 & 15587.09 \\ 
    & distance to nearest paper mills and products site (m) & 24947.87 & 33794.48 \\ 
    & distance to nearest paints and coatings site (m) & 11514.88 & 17416.26 \\ 
    & distance to nearest national defense site (m) & 22456.74 & 27376.78 \\ 
    & distance to nearest mining and refining site (m) & 23174.41 & 22574.16 \\ 
    & distance to nearest metal machinery manufacturing site (m) & 9813.88 & 18371.10 \\ 
    & distance to nearest metal coating site (m) & 7558.19 & 12846.41 \\ 
    & distance to nearest industrial gas site (m) & 24988.08 & 31728.17 \\ 
    & distance to nearest glass products site (m) & 26884.94 & 33812.39 \\ 
    & distance to nearest fire training site (m) & 37445.49 & 34881.62 \\ 
    & distance to nearest electronics industry site (m) & 9150.41 & 15660.93 \\ 
    & distance to nearest consumer products site (m) & 15449.11 & 20938.36 \\ 
    & distance to nearest cleaning product manufacturing site (m) & 14065.07 & 19240.38 \\ 
    & distance to nearest chemical manufacturing site (m) & 8812.54 & 13219.36 \\ 
    & distance to nearest cement manufacturing site (m) & 27534.49 & 31156.89 \\ 
    & distance to nearest airport (part 139) (m) & 23247.07 & 21401.99 \\ 
    & distance to nearest airport (m) & 9707.06 & 12254.24 \\ 
    \hline
  CalEnviroScreen & CalEnviroScreen Score, Pollution Score multiplied by Population Characteristics Score & 33.04 & 15.90 \\ 
    & Percentile of the CalEnviroScreen score, grouped by 5\% increments & 58.91 & 27.05 \\ 
    & Amount of daily maximum 8 hour Ozone concentration & 0.05 & 0.01 \\ 
    & Ozone percentile & 58.86 & 27.49 \\ 
    & Annual mean PM2.5 concentrations & 10.18 & 2.71 \\ 
    & PM2.5 percentile & 49.89 & 32.95 \\ 
    & Diesel PM emissions from on-road and non-road sources & 0.20 & 0.19 \\ 
    & Diesel PM percentile & 47.46 & 30.15 \\ 
    & Drinking water contaminant index for selected contaminants & 579.20 & 204.01 \\ 
    & Drinking water percentile & 63.72 & 24.24 \\ 
    & Potential risk for lead exposure in children living in low-income communities with older housing & 48.10 & 22.03 \\ 
    & Children's lead risk from housing percentile & 48.87 & 28.09 \\ 
    & Total pounds of selected active pesticide ingredients (filtered for hazard and volatility) used in production-agriculture per square mile & 1049.65 & 5341.53 \\ 
    & Pesticides percentile & 32.71 & 36.79 \\ 
    & Toxicity-weighted concentrations of modeled chemical releases to air from facility emissions and off-site incineration (from RSEI) & 1426.31 & 2925.34 \\ 
    & Toxic release percentile & 43.10 & 31.41 \\ 
    & Traffic density in vehicle-kilometers per hour per road length, within 150 meters of the census tract boundary & 1128.36 & 868.48 \\ 
    & Traffic percentile & 49.83 & 30.44 \\ 
    & Sum of weighted EnviroStor cleanup sites within buffered distances to populated blocks of census tracts & 15.05 & 25.48 \\ 
    & Cleanup sites percentile & 45.76 & 35.67 \\ 
    & Sum of weighted GeoTracker leaking underground storage tank sites within buffered distances to populated blocks of census tracts & 28.42 & 47.78 \\ 
    & Groundwater threats percentile & 48.11 & 34.49 \\ 
    & Sum of weighted hazardous waste facilities and large quantity generators within buffered distances to populated blocks of census tracts & 1.19 & 2.32 \\ 
    & Hazardous waste percentile & 57.33 & 30.63 \\ 
    & Sum of number of pollutants across all impaired water bodies within buffered distances to populated blocks of census tracts & 4.10 & 5.27 \\ 
    & Impaired water bodies percentile & 33.78 & 32.76 \\ 
    & Sum of weighted solid waste sites and facilities (SWIS) within buffered distances to populated blocks of census tracts & 4.25 & 5.96 \\ 
    & Solid waste percentile & 45.38 & 36.98 \\ 
    & Average of percentiles from the Pollution Burden indicators (with a half weighting for the Environmental Effects indicators) & 48.23 & 13.29 \\ 
    & Pollution Burden variable scaled with a range of 0-10. (Used to calculate CES 4.0 Score) & 5.89 & 1.62 \\ 
    & Pollution burden percentile & 61.24 & 28.26 \\ 
    & Age-adjusted rate of emergency department visits for asthma & 52.90 & 27.08 \\ 
    & Asthma percentile & 52.58 & 26.02 \\ 
    & Percent low birth weight & 4.95 & 1.40 \\ 
    & Low birth weight percentile & 49.46 & 27.09 \\ 
    & Age-adjusted rate of emergency department visits for heart attacks per 10,000 & 14.14 & 5.02 \\ 
    & Cardiovascular disease percentile & 54.76 & 27.59 \\ 
    & \% population over 25 with less than a high school education & 19.85 & 13.64 \\ 
    & Education percentile & 56.87 & 25.63 \\ 
    & Percent limited English speaking households & 9.32 & 8.61 \\ 
    & Linguistic isolation percentile & 48.19 & 28.58 \\ 
    & \% population living below two times the federal poverty level & 35.03 & 16.76 \\ 
    & Poverty percentile & 57.12 & 25.94 \\ 
    & \% the population over the age of 16 that is unemployed and eligible for the labor force & 6.66 & 3.80 \\ 
    & Unemployment percentile & 54.17 & 29.19 \\ 
    & Percent housing-burdened low-income households & 17.49 & 7.54 \\ 
    & Housing burden percentile & 47.13 & 28.02 \\ 
    & Average of percentiles from the Population Characteristics indicators & 52.48 & 18.48 \\ 
    & Population Characteristics variable scaled with a range of 0-10. (Used to calculate CES 4.0 Score) & 5.44 & 1.92 \\ 
    & Population characteristics percentile & 53.77 & 26.24 \\ 
    \hline
  USGS & elevation at well site (meters above sea level) & 152.27 & 240.75 \\ 
  & soil silt content at well site (\%) & 27.52 & 11.18 \\ 
  & soil sand content at well site (\%) & 54.63 & 19.84 \\ 
  & soil clay content at well site (\%) & 15.38 & 11.27 \\ 
  & soil organic matter content at well site (log10(\%)) & 1.36 & 1.61 \\ 
  & soil bulk density at well site (grams per cubic centimeter) & 1.49 & 0.09 \\ 
  & saturated soil hydraulic conductivity at well site (log10(centimters per hour)) & 6.46 & 6.98 \\ 
  & soil pH at well site & 6.83 & 0.68 \\ 
   \hline
   \label{tab:app_covdes}
\end{longtable}

\bibliographystyle{elsarticle-harv} 
\bibliography{refs}

\end{document}